\documentclass[preprint]{aastex}
\usepackage{amssymb, amsmath, rotating}
\usepackage{arydshln}
\usepackage{color}

\newcommand{\dt}{\delta t}

\newcommand{\changetext}[1]{{\textcolor{black}{#1}}}
\newcommand{\suma}{{\Sigma_a}}
\newcommand{\bfea}{a_{\Delta i \Delta j}}

\begin{document}

\title{Brighter-fatter effect in near-infrared detectors -- I. Theory of flat auto-correlations}

\date{\today}
\author{Christopher M. Hirata}
\email{hirata.10@osu.edu}
\author{Ami Choi}
\affil{Center for Cosmology and AstroParticle Physics, The Ohio State University, 191 West Woodruff Avenue, Columbus, Ohio 43210, USA}

\begin{abstract}
Weak gravitational lensing studies aim to measure small distortions in the shapes of distant galaxies, thus placing very tight demands on the understanding of detector-induced systematic effects in astronomical images. The Wide-Field Infrared Survey Telescope (WFIRST) will carry out weak lensing measurements in the near infrared using the new Teledyne H4RG-10 detector arrays, which makes the range of possible detector systematics very different from traditional weak lensing measurements using optical CCDs. One of the non-linear detector effects observed in CCDs is the brighter-fatter effect (BFE), in which charge already accumulated in a pixel alters the electric field geometry and causes new charge to be deflected away from brighter pixels. Here we describe the formalism for measuring the BFE using flat field correlation functions in infrared detector arrays. The auto-correlation of CCD flat fields is often used to measure the BFE, but because the infrared detector arrays are read out with the charge ``in place,'' the flat field correlations are dominated by capacitive cross-talk between neighboring pixels (the inter-pixel capacitance, or IPC).  \changetext{Conversely, if the BFE is present and one does not account for it, it can bias correlation measurements of the IPC and photon transfer curve measurements of the gain.} We show that one can compute numerous cross-correlation functions between different time slices of the same flat exposures, and that correlations due to IPC and BFE leave distinct imprints. We generate a suite of simulated flat fields and show that the underlying IPC and BFE parameters can be extracted, even when both are present in the simulation. There are some biases in the BFE coefficients up to 12\%, which are likely caused by higher order terms that are dropped from this analysis. The method is applied to laboratory data in the companion Paper II.
\end{abstract}


\section{Introduction}
\label{sec:intro}

Weak gravitational lensing (WL) -- the distortion of the shapes and sizes of distant galaxies by the curvature of the intervening space-time -- is a powerful method for probing the matter distribution in the Universe (for recent results, see e.g.\ \citealt{2013MNRAS.432.2433H}; \citealt{2016PhRvD..94b2001A}; and \citealt{2017MNRAS.465.1454H}). However, the signal is small and must be measured to a fraction of a percent to meet the science goals of current and future WL surveys. Therefore, WL programs place a very strong emphasis on understanding every systematic error that can occur in measuring the shape of the galaxy -- this includes the contribution of the atmosphere, optics, and image motion to the smearing of an image, as well as imprints of the detector system and data processing.

The brighter-fatter effect (BFE; e.g.\ \citealt{2014JInst...9C3048A}) is one of these subtle effects that has been observed in silicon CCD detectors. This is a non-linear effect in which a brighter point source produces a larger image (as measured by e.g.\ full width at half maximum) in the CCD than a fainter point source. It is caused by changes in the electric field geometry in the CCD as a well fills up with electrons: if at any instant during the exposure a pixel $(i,j)$ contains more charge than its neighbors, then due to self-repulsion of the electrons, additional photo-electrons generated will be less likely to land in pixel $(i,j)$ and more likely to land in its neighbors. This is described phenomenologically for CCDs by supposing that the pixel boundaries\footnote{A ``pixel boundary'' in an either astronomical CCD or a NIR detector is not a physical barrier between neighboring pixels, but rather is defined by the electric fields and diffusion coefficients that determine which well ultimately collects a photo-electron.} move as a function of accumulated charge. In thick CCDs, the BFE has been observed to have a significant range, e.g.\ in the Dark Energy Camera (DECam) CCDs, the pixel boundary shifts have been measured from charge up to $\sim 10$ pixels away \citep{2015JInst..10C5032G}. The BFE also manifests itself in correlation properties (variance and correlation function) of flat-field images, where the shifting pixel boundaries break the usual assumption that each photo-electron behaves independently from previous electrons and hence causes non-Poisson correlations in the flat images \citep{2015A&A...575A..41G}. Indeed, this provided one of the early hints to the existence of the BFE \citep{2006SPIE.6276E..09D}.

The BFE and techniques for modeling it have been well-established in current WL surveys such as the Dark Energy Survey (DES; \citealt{2015JInst..10C5032G}) and the Hyper-Suprime Cam (HSC; \citealt{2018AJ....155..258C}). Higher precision will be demanded of the next generation of WL surveys: the Large Synoptic Survey Telescope (LSST), the Euclid space mission, and the Wide-Field Infrared Survey Telescope (WFIRST) space mission. The BFE has been observed and characterized in Euclid development CCDs \citep{2015ExA....39..207N} and in candidate sensors for LSST \citep{2015JInst..10C5024B, 2017JInst..12C3091L}. In setting requirements on any detector effect, it is important to study all of the ways that effect can enter into the analysis; in the case of the brighter-fatter effect in weak lensing, the stars used for determination of the point spread function (PSF) are much brighter (and have steeper intensity gradients) than either the galaxies or the sky background, and so the stars are most affected by the BFE. This means that the principal effect of the BFE is to make the measured (star-based) PSF larger than the correct PSF for faint galaxies. Subsequent stages of analysis will then over-correct for the smearing out of galaxy ellipticities by the finite size of the PSF, and hence will over-estimate the shear signal. Note that we consider the BFE to be a calibration problem, in the sense that WFIRST will need to develop a model for it; we do not need to eliminate it.

WFIRST plans to measure weak lensing in the near infrared (NIR), and thus silicon CCDs are not an option. Instead, it will use \changetext{infrared detector arrays}: each of the 18 detector arrays will use Teledyne's H4RG-10 readout integrated circuit ($4088\times 4088$ active pixels, 10 $\mu$m pitch) with mercury cadmium telluride (HgCdTe; $2.5$ $\mu$m cutoff) as the light-absorbing component.\footnote{For more background on the H$x$RG series devices, we refer the reader to the overview \citep{2008SPIE.7021E..0HB, 2011ASPC..437..383B}, and reports on the JWST/NIRSpec H2RGs \citep{2007PASP..119..768R, 2014PASP..126..739R} and on the WFIRST H4RG development \citep{2014SPIE.9154E..2HP}.} Just as for CCDs, the boundaries between adjacent pixels are defined by the solution to the drift-diffusion equation rather than a physical barrier, and so a BFE in WFIRST detectors would be physically plausible; however one would expect the details to be very different. It is therefore important to understand whether the BFE, or other non-linear detector effects, are present in WFIRST prototype devices, and if so how the BFE fits into the overall WFIRST calibration plan. A discussion of the physics of the BFE as applied to NIR detectors, as well as some characterization efforts on earlier generations of Teledyne detectors (H1RG and H2RG, 18 $\mu$m pitch) can be found in \citet{2017JInst..12C4009P}. The BFE has been measured in point source illumination tests on an H2RG tested for the {\slshape Euclid} program \citep{2018PASP..130f5004P}.

In order to develop calibration and data reduction procedures for WFIRST data, it is essential to characterize non-linear effects such as the BFE. Such characterization efforts also provide us with the opportunity to learn about what calibration procedures work and what pitfalls exist. Such studies must be undertaken early in the life cycle of the project, especially in the context of a space mission where late changes to the calibration requirements could be expensive or impossible. Detailed characterization of subtle effects such as the BFE is also outside the scope of the technology development milestones \citep{2015arXiv150303757S}. These milestones were recently completed for the WFIRST NIR detectors, but focus on basic performance (e.g.\ quantum efficiency, read noise, persistence), production yield, and environmental testing (e.g.\ thermal cycling, vibrational, radiation). However, the studies for these milestones do not explore the detector-related systematic effects that are relevant to control shear errors at the few$\,\times 10^{-4}$ level.

The two major methods of measuring the BFE are (i) measurement of spots projected onto the detector (either with a laboratory spot projector or using real stars observed through a telescope) and (ii) flat field statistics. This paper considers flat fields, since spots were not yet available for the H4RG-10 HgCdTe detectors at the time we began this project.\footnote{Jay Anderson (private communication) has presented some results on the BFE in the HST WFC3-IR channel using observations of stars to the WFIRST Detector Working Group. While WFC3-IR is a useful guide to some of the issues WFIRST will encounter, it is an 18 $\mu$m pitch device and so it is important to measure BFE parameters on H4RG-10 HgCdTe devices.} The data were acquired as part of general detector characterization and made available to the science teams; the test procedure was not specifically optimized for BFE studies.

The use of flat field statistics for BFE measurements presents some special challenges \changetext{for infrared arrays}, most notably that the flat field auto-correlation function is dominated by the effect of inter-pixel capacitance (IPC), which gives a positive correlation between adjacent pixels \citep{2004SPIE.5167..204M}. IPC can have a linear component and a non-linear component, the latter of which is known as non-linear inter-pixel capacitance (NL-IPC, a signal dependent coupling that occurs when one converts from charge to voltage; e.g., \citealt{Cheng2009, 2016SPIE.9915E..2ID, 2017OptEn..56b4103D, 2018PASP..130g4503D}).\footnote{There are also reports of NL-IPC in H2RG detectors (Arielle Bertrou-Cantou, private communication).} On the other hand, the H4RG-10 provides a non-destructive read capability, which is very useful for BFE studies as it enables intermediate stages of the image to be observed: the flat field is then a 3D data cube, and correlations across different time slices (``frames'') of the image can be measured. We will find that this capability allows us to simultaneously measure the IPC and BFE using flat fields. We will further find that the gain measurement from the photon transfer curve \citep[e.g.,][]{1981SPIE..290...28M, 1985SPIE..570....7J} must be corrected for the BFE in addition to the now-standard IPC correction \citep{2006OptEn..45g6402M, 2006PASP..118.1443B}.

The cleanest method for BFE measurements from H4RG-10 flat fields is to cross-correlate two correlated double sample (CDS) images, obtained from {\em non-overlapping} parts of the ramp. This eliminates any possible correlations from Poisson noise -- including those that couple between neighboring pixels through linear IPC -- as well as any read noise correlations that occur within a single frame. It does leave correlations due to classical non-linearity, which must be removed based on the standard non-linearity curve analysis. The non-overlapping correlation function method cannot tell the difference between the brighter-fatter effect (which occurs during the process of collecting charge into a well) from NL-IPC.  To distinguish the BFE from NL-IPC, we must resort to correlations of CDS images over the {\em same} (or at least an overlapping) time interval, and observe how pixel variances or covariances of adjacent pixels change as one varies the time interval of interest. While these tests mix together many different detector effects, the BFE and NL-IPC hypotheses make distinct predictions. NL-IPC appears in these methods with a factor of 2 different from BFE, because in the BFE each electron\footnote{\changetext{The Teledyne HxRG detectors actually collect holes, rather than electrons, although the statistical techniques in this paper are agnostic to the sign of the charge. We will use the standard nomenclature of charge in ``electrons'' -- this is common usage in the astronomical community and HST WFC3-IR documentation, even though the WFC3-IR detector also collects holes.}} collected only affects the behavior of subsequent electrons, but in NL-IPC (which acts on collected charge) every electron affects every other electron.\footnote{In \S\ref{sec:theory}, the mathematics of this is worked out in great detail, but the underlying reason for the factor of 2 is the simple combinatorial effect.}

This paper is organized as follows. In \S\ref{sec:formalism}, we build up our description of the brighter-fatter effect as well as other detector effects relevant to the flat field (IPC and non-linearity) and the formalism for correlation functions among the frames of a flat field. In \S\ref{sec:theory}, we work out the theoretical predictions for the 2-point correlation function of the flat fields in the presence of the various effects. In \S\ref{sec:sim}, we describe a simulation incorporating IPC, non-linearity, and the BFE that allows us to test the characterization methods in \S\ref{sec:char}. We conclude in \S\ref{sec:discussion}. The appendix contains some technical material on the covariances of clipped data (Appendix~\ref{app:covar-clip}). The application to laboratory H4RG-10 data -- and associated evidence for the BFE -- is presented in a companion paper (Paper II).

\section{Formalism}
\label{sec:formalism}

\subsection{Brighter-fatter effect}

Autocorrelation measurements are sensitive to the BFE via changes in the effective pixel area.\footnote{In the BFE literature, the operational definition of pixel ``area'' is that ${\cal A}_{i,j} = {\rm QE}_{\rm ref}^{-1} \int p_{i,j}(x,y)\,dx\,dy$, where $p_{i,j}(x,y)$ is the probability that a photon incident at position $(x,y)$ on the detector leads to an electron collected in pixel $(i,j)$. This integral is divided by a reference value of the quantum efficiency, QE$_{\rm ref}$, so that at low flux levels the sum of pixel areas in some region corresponds to the geometrical area. If the probability of collecting a charge in {\em any} well $\sum_{i,j} p_{i,j}(x,y)$ remains fixed, then the BFE conserves total pixel area.} We suppose that a pixel $(i,j)$ has effective area that changes depending on the charge in neighboring pixels:
\begin{equation}
{\cal A}_{i,j} = {\cal A}_{i,j}^0 \left[ 1 + \sum_{\Delta i,\Delta j} a_{\Delta i,\Delta j} Q({i+\Delta i,j+\Delta j}) \right],
\label{eq:Aij}
\end{equation}
where $i$ and $j$ denote column and row indices (0...4095 for the H4RG), $Q({i,j})$ is the charge (in number of elementary charges) in pixel $(i,j)$, and $a_{\Delta i,\Delta j}$ denotes a coupling matrix. In a general case, one might allow $a$ to also depend on $i$ and $j$, which would correspond to a BFE that varies from one pixel to another. However with flat autocorrelation data and a plausible number of flats, it is possible only to measure averages of the $a$ coefficients in groups of pixels. Therefore we will assume discrete translation invariance; note, however, that the ability to perform autocorrelations with different sets of pixels allows some (limited) ability to test for translation invariance.

While $a_{\Delta i,\Delta j}$ is formally dimensionless, we will normally quote $a_{\Delta i,\Delta j}$ in units of 10$^{-6}$ e$^{-1}$, ppm/e, or \%/$10^4$ e (all of which are equivalent). These units are convenient because $10^4$ e is a typical integrated signal level in the central pixel of a PSF star for WFIRST, so a measured value of $a$ in ppm/e maps into the expected order of magnitude of the effect on a star in percent.

In a phenomenological BFE model, one specifies how much of the area change comes from each of the boundaries by writing
\begin{equation}
a_{\Delta i,\Delta j} = a_{\Delta i,\Delta j}^R + a_{\Delta i,\Delta j}^T + a_{\Delta i,\Delta j}^L + a_{\Delta i,\Delta j}^B,
\end{equation}
where the superscripts $R$, $T$, $L$, and $B$ refer to the right, top, left, and bottom boundaries respectively. If the quantum efficiency depends on the charge in the well, then we would include an additional term $a_{\Delta i,\Delta j}^{QE}$. An image simulation requires as input all of these components individually, and they can be probed with spot illumination or individual pixel resets; however flat autocorrelations are only sensitive to the total.

A true BFE that works by moving pixel boundaries conserves total area, so we should have
\begin{equation}
\sum_{\Delta i,\Delta j} a_{\Delta i,\Delta j} = 0.
\label{eq:BFE-SumRule}
\end{equation}
(We expect that $a_{0,0}$ would be negative, and the $a$ coefficients for the neighbors would be positive.)
However, the sum in Eq.~(\ref{eq:BFE-SumRule}) is ill-behaved, since the noise diverges as we continue to add pixels. It can therefore be tested only in the context of fitting a model to $a_{\Delta i,\Delta j}$. Moreover, if adding charge to a pixel changes the QE or charge collection probability, then Eq.~(\ref{eq:BFE-SumRule}) may be violated. Therefore, it is important to measure all of the $a_{\Delta i,\Delta j}$, without assuming Eq.~(\ref{eq:BFE-SumRule}). In general we will define:
\begin{equation}
\suma = \sum_{\Delta i,\Delta j} a_{\Delta i,\Delta j}.
\end{equation}
We will find it useful later to define
\begin{equation}
a'_{\Delta i,\Delta j} \equiv a_{\Delta i,\Delta j} - \delta_{\Delta i,0}\delta_{\Delta j,0} \suma.
\end{equation}
By definition, the $a'$ coefficients sum to zero. Some of the BFE tests that we will conduct are not sensitive to $\suma$, and hence can only measure the $a'_{\Delta i,\Delta j}$.

Finally, we define symmetry-averaged functions over coefficients $(\Delta i,\Delta j)$ related by the rotation or reflection symmetries of the pixel grid:
\begin{equation}
a_{\langle \Delta i,\Delta j \rangle} \equiv
\frac18\Bigl(
a_{\Delta i,\Delta j} + a_{\Delta j,\Delta i} + a_{-\Delta j,\Delta i} + a_{-\Delta i,\Delta j}
+ a_{-\Delta i,-\Delta j} + a_{-\Delta j,-\Delta i} + a_{\Delta j,-\Delta i} + a_{\Delta i,-\Delta j}
\Bigr).
\label{eq:gamma-ave}
\end{equation}
There is no law of physics requiring the BFE to respect the rotation and reflection symmetries (indeed, in CCDs it does not), so some test results are provided for, e.g.\ $a_{\Delta i,\Delta j}$ and $a_{\Delta j,\Delta i}$ separately.

\subsection{Gains, nonlinearities, and IPC}

``Raw'' data from the detector arrays are not in electrons but in data numbers (DN), which are voltages quantized as 16-bit integers. As each pixel is exposed to light, the voltage \changetext{difference} across the photodiode decreases; \changetext{for detectors that collect holes, the voltage on the readout (p-type) side of the diode increases. The observed signal $S$ (units: DN) may increase or decrease depending on the polarity of the analog-to-digital converter.\footnote{\changetext{We have worked with raw data of both polarities, and it is easy enough to switch between them by inserting an optional mapping $S\rightarrow 2^{16}-1-S$ in the input script.}} In this paper we work with the convention that $S$ decreases during integration.} Ideally the relation between the accumulated charge and signal drop would be linear, but in practice it is not. This effect can contain contributions both from the non-linearity of the $p-n$ junction itself as well as any step in the readout chain, and is generically observed in NIR detectors \citep[e.g.][]{2005nicm.rept....2B,2010SPIE.7731E..3CD,2010SPIE.7742E..22H}; see \citet{2016PASP..128j4001P} for a study of its impact on the WFIRST weak lensing program.

A polynomial model is typically used to describe the non-linearity curve; the most important correction is typically the quadratic term. In flat illumination, where each pixel accumulates charge $Q$, the drop in signal level is given by
\begin{equation}
S_{\rm initial} - S_{\rm final} = \frac1g\left( Q - \beta Q^2 \right),
\label{eq:S-g-beta}
\end{equation}
where $g$ is the gain (units: e/DN) and $\beta$ is the leading-order non-linearity coefficient. Note that $\beta$ has the same units as ${a}$, and so it will be convenient to quote it in ppm/e. We define ``initial'' for the purposes of Eq.~(\ref{eq:S-g-beta}) to mean immediately following a reset, which we take to be $t=0$.  Note that non-linearity parameters will likely depend on the reset voltage.

In thick CCD detectors, it is common to use auto-correlations of the flat field to measure the BFE. However, in \changetext{infrared} detectors the auto-correlation of a flat is instead dominated by inter-pixel capacitance (IPC). IPC is an electrical coupling between neighboring pixels, in which the voltage on one pixel is sensitive to the charge in its neighbors \citep[e.g.][]{2004SPIE.5167..204M, 2006OptEn..45g6402M}. This coupling increases the apparent size of the image of a star on the detector; see \citet{2016PASP..128i5001K} for a study of its impact on WFIRST.
This means a more complex procedure is needed to probe the BFE in \changetext{infrared arrays}. Furthermore, determination of the gain from variance vs.\ mean plots must be corrected for IPC to obtain meaningful results \citep[e.g.][]{2006OptEn..45g6402M, 2008SPIE.7021E..23F, 2012SPIE.8453E..1RC}. Normally the IPC is described by replacing the linear term in Eq.~(\ref{eq:S-g-beta}) with a kernel describing the capacitive cross-talk among the pixels:
\begin{equation}
S_{\rm initial}(i,j) - S_{\rm final}(i,j) = \frac1g \sum_{\Delta i,\Delta j} K_{\Delta i,\Delta j} Q_{i-\Delta i, j-\Delta j}
+ [{\rm nonlinear~terms}]
,
\end{equation}
where the kernel matrix ${\bf K}$ satisfies
\begin{equation}
\sum_{\Delta i,\Delta j} K_{\Delta i,\Delta j} = 1.
\label{eq:norm-K}
\end{equation}
In the case where the IPC only talks to the nearest neighbors and does so equally, we have $K_{0,0} = 1-4\alpha$, $K_{0,\pm 1} = K_{\pm 1,0} = \alpha$, and all others are zero. However, asymmetries between the horizontal and vertical directions ($K_{0,\pm 1}\neq K_{\pm 1,0}$) are commonly observed in NIR detectors. Therefore we measure separately $\alpha_{\rm H} = K_{\pm 1,0}$ and $\alpha_{\rm V} = K_{0,\pm 1}$; if these are different then we define $\alpha$ to be their average $(\alpha_{\rm H}+\alpha_{\rm V})/2$. We will also allow for diagonal IPC, $\alpha_{\rm D} = K_{\pm 1, \pm 1}$ (when this notation is used, we will not distinguish between the ``northeast-southwest'' and ``northwest-southeast'' directions, although in principle their IPC may be different).

Inter-pixel capacitance in a semiconductor device may depend on signal since such devices do not obey the principle of superposition. This ``non-linear inter-pixel capacitance'' (NL-IPC) can be phenomenologically similar to the BFE: if the IPC grows with signal level, then this will also lead to brighter stars showing a larger observed FWHM on account of the greater amount of coupling. However, NL-IPC is a different mechanism -- it arises in the conversion of charge to voltage, whereas the BFE arises in the collection of charge -- and as such there are subtle differences in how it impacts both flat field statistics and science data. Disentangling the two effects proves to be one of the most difficult part of the BFE analysis.

Discussions of NL-IPC are complicated by the fact that NIR detectors both have non-linear charge-to-signal conversion (Eq.~\ref{eq:S-g-beta}) and IPC. In the presence of both of these effects, we generically expect some kind of non-linear cross-talk between neighboring pixels of order $\alpha\beta$, and great care is needed to even define a quantitative measure of NL-IPC. In most astronomical data processing, the non-linearity correction is applied to individual pixels as one of the first steps -- and certainly before any attempt at IPC correction (if the latter is done at all). This is equivalent to the assumption that {\em all} of the non-linearity acts on the signal {\em after} IPC. In this paper, we use ``NL-IPC'' to denote any non-linearity in the charge-to-signal conversion that deviates from this assumption.

For the purposes of flat fields, we parameterize NL-IPC by a mean signal-level-dependent kernel,
\begin{equation}
S_{\rm initial}(i,j) - S_{\rm final}(i,j) = \frac1g \sum_{\Delta i,\Delta j} [K_{\Delta i,\Delta j} + K'_{\Delta i,\Delta j} \bar Q] Q_{i-\Delta i, j-\Delta j},
\label{eq:NL-IPC-general}
\end{equation}
where $\bar Q$ is the mean accumulated charge ($It$ in a flat exposure). One can equivalently write this in terms of $\alpha'$, $\alpha'_{\rm H}$, $\alpha'_{\rm V}$, etc.:
\begin{eqnarray}
S_{\rm initial}(i,j) - S_{\rm final}(i,j) &=& \frac1g \Bigl[
Q_{i,j} + (\alpha_{\rm H}+\alpha'_{\rm H}\bar Q) (Q_{i+1,j}-Q_{i,j})
 + (\alpha_{\rm H}+\alpha'_{\rm H}\bar Q) (Q_{i-1,j}-Q_{i,j})
\nonumber \\ &&
  + (\alpha_{\rm V}+\alpha'_{\rm V}\bar Q) (Q_{i,j+1}-Q_{i,j})
   + (\alpha_{\rm V}+\alpha'_{\rm V}\bar Q) (Q_{i,j-1}-Q_{i,j})
\Bigr].
\label{eq:NL-IPC-alpha}
\end{eqnarray}
In the flat illumination case, there is no ambiguity of what mean accumulated charge $\bar Q$ should be used. In other cases such as spot illumination or pixel reset tests, neighboring pixels can have wells filled to very different levels and $K'$ is no longer the appropriate concept. Some studies have indicated that NL-IPC is a function of both contrast and signal level -- see, e.g., \citet{2016SPIE.9915E..2ID, 2017OptEn..56b4103D, 2018PASP..130g4503D} -- and in this case one should write an IPC coupling constant $\alpha(Q_{i,j},Q_{i+1,j})$ that is a function of charge in both pixels.\footnote{If viewed as a capacitor network, voltage in the pixel might be a more fundamental variable than the charge.} The flat field test probes the case of $Q_{i+1,j}\approx Q_{i,j} \approx \bar Q$, whereas single pixel reset and hot pixel tests measure the case where $Q_{i+1,j}\approx 0$.

\subsection{Correlation functions}

In a CCD flat, there is only a single read of the detector in each flat exposure. However, in an \changetext{infrared array} flat, one typically obtains $N$ samples up the ramp, and correlation functions can be defined not just between different pixels but between different frames. If one denotes the frames by indices $abc...$ then let us define:
\begin{equation}
C_{abcd}(\Delta i,\Delta j) = {\rm Cov} \left[
S_a(i,j) - S_b(i,j), S_c(i+\Delta i,j+\Delta j) - S_d(i+\Delta i,j+\Delta j)
\right].
\label{eq:corrfunc}
\end{equation}
If one obtained $N=2$ samples and took the autocorrelation of the CDS image $S_1-S_2$, this would correspond to $C_{1212}(\Delta i,\Delta j)$. This is the procedure that is most similar to a CCD flat autocorrelation. However, as noted above, it contains IPC as well as BFE and therefore cannot distinguish the two. Fortunately, with multiple up-the-ramp samples \changetext{an infrared array} flat is much richer in information than a CCD flat, and the {\em temporal} structure ($abcd$ indices) is the key to disentangling the various effects.

In what follows, we will simplify some expressions by writing $S_{ab}(i,j) \equiv S_a(i,j) - S_b(i,j)$. (Note the sign convention!)

The correlation functions satisfy the trivial properties:
\begin{list}{$\bullet$}{}
\item $C_{abcd}(\Delta i,\Delta j)=0$ if $a=b$ or $c=d$.
\item $C_{abcd}(\Delta i,\Delta j)=C_{cdab}(-\Delta i,-\Delta j)$.
\item $C_{abcd}(\Delta i,\Delta j)=-C_{bacd}(\Delta i,\Delta j)=-C_{abdc}(\Delta i,\Delta j)=C_{badc}(\Delta i,\Delta j)$.
\item $C_{abcd}(\Delta i,\Delta j) = C_{aecd}(\Delta i,\Delta j)+C_{ebcd}(\Delta i,\Delta j)$
and
$C_{abcd}(\Delta i,\Delta j) = C_{abcf}(\Delta i,\Delta j)+C_{abfd}(\Delta i,\Delta j)$.
\end{list}
The last property means that all of the correlation functions can be composed of ``elementary'' correlation functions $C_{a,a+1,c,c+1}(\Delta i,\Delta j)$.

Equation~(\ref{eq:corrfunc}), like all covariance matrices, requires more than one realization to make a measurement. The standard approach, followed here, is to compare a pair of two flats. The ``measured covariance'' of any two observables ${\cal O}$ and ${\cal O}'$ is then
\begin{equation}
{\rm Cov}_{\rm meas}\left[ {\cal O},{\cal O}' \right] =
\frac12 \langle
({\cal O}_A-{\cal O}_B)
({\cal O}'_A-{\cal O}'_B)
\rangle,
\end{equation}
where the average is taken over pixels $(i,j)$ in the region of interest. The differencing removes small deviations such as imperfect illumination patterns, intrinsic variations in pixel area or QE, etc.

In a flat or dark field with many samples, we may also construct a correlation function averaged in the time direction. If we take the average of $n$ time-translations of the time windows $abcd$, then we find
\begin{equation}
\bar C_{abcd[n]}(\Delta i,\Delta j) = \sum_{\nu=0}^{n-1}
C_{a+\nu,b+\nu,c+\nu,d+\nu}(\Delta i,\Delta j).
\label{eq:corrfunc-ave}
\end{equation}
Since a flat field is not time-stationary (gain, non-linearity, and possibly other quantities will change as the voltage across the $p-n$ junction decreases), we must keep track of all the time indices $abcdn$ when fitting a model to the time-translation-averaged correlation function.

While in this study we consider individual time samples, space-based infrared surveys are often data rate limited and thus not every sample can be downlinked. Therefore future work should also examine how the BFE and other effects appear in the cross-correlation functions of flat field data with compression along the time axis, e.g., using the first few Legendre coefficients \citep{2019arXiv190202312R}. If compression by linear combinations is used (Legendre coefficients, group averaging, etc.), any such cross-correlation function can be written trivially as an appropriate weighted sum over the $C_{abcd}(\Delta i,\Delta j)$.

\subsection{Multiple exposures}

When we discuss statistical algorithms, it will be essential to describe operations acting on multiple exposures. Here a specific exposure will be denoted with a $|$ separator, followed by the exposure type and number. For example, we write $S_a(i,j|{\rm F}_k)$ to denote the signal in time step $a$ and pixel $(i,j)$ in the $k^{\rm th}$ flat field, and $S_a(i,j|{\rm D}_k)$ for the $k^{\rm th}$ dark exposure. This formalism could be extended in the future to include other types of tests (besides flats and darks).

\section{Theory}
\label{sec:theory}

\changetext{We now embark on the main calculation in this paper: the determination of the correlation function $C_{abcd}(\Delta i,\Delta j)$ including IPC, classical non-linearity, and the brighter-fatter effect to leading order. We will summarize the general result in Eq.~(\ref{eq:master-cf}). The measured correlation functions can then be used to simultaneously constrain the gain, IPC, and BFE parameters of a detector array or sub-region thereof -- a step we will take on real data in Paper II. We will assume $a<b$ and $c<d$, since these functions contain all the information because of symmetries, but we do not assume anything else about the ordering. In particular, the exposure intervals $a...b$ and $c...d$ may be the same, may overlap, or may be non-overlapping. The ``same interval'' case $(a,b)=(c,d)$ will be the most familiar to readers who have worked with 2D image products before (e.g., CCD images, or CDS images from infrared arrays). However, the non-overlapping case turns out to be of particular use for measuring the sum of BFE and NL-IPC. It is the combination of these many different cases (itself possible due to the non-destructive read capability) that allows us to constrain so many parameters.}

\changetext{Rather than trying to solve everything at once, we begin this section by considering a ``perfect'' detector (no IPC, BFE, or any form of non-linearity; \S\ref{ss:D1}), and then adding layers of physical and mathematical complexity. In particular, we add linear inter-pixel capacitance, which simply introduces a convolution kernel (\S\ref{ss:D2}). Next is the classical non-linearity (\S\ref{ss:D3}), where covariances of higher-order moments of the charge appear; then we put the classical non-linearity together with the IPC (\S\ref{ss:D4}). We introduce the brighter-fatter effect in \S\ref{ss:D5}; there the buildup of charge in each pixel $Q(i,j;t)$ is a stochastic process with interactions between pixels, and we write and solve a differential equation for the moments of the charge. Non-linear IPC is briefly discussed and included in \S\ref{ss:D6}. The main result, Eq.~(\ref{eq:master-cf}), is presented in \S\ref{ss:combined-cf}, and we discuss some special cases in \S\ref{ss:specialcases}. A reader not interested in the details of the derivation may skip directly to Eq.~(\ref{eq:master-cf}) and the simplifications in \S\ref{ss:specialcases} (though one will still have to refer to the definitions for the quantities that summarize the time intervals: $t_{abcd}$, $T_{abcd}$, $\sigma_{abcd}$, and $\tau_{abcd}$, defined in Eqs.~\ref{eq:tabcd-def}, \ref{eq:_Tabcd-def}, \ref{eq:sigma-abcd-def}, and \ref{eq:tau-abcd-def}, respectively).}

\changetext{In this calculation,} we suppose that the flat illumination provides current $I$ per pixel (units: e/s) and that the frame $a$ is saved at time $t_a$. We \changetext{also} assume that the flat field illumination uses a wavelength long enough for quantum yield effects to be insignificant (i.e., where one photon produces at most one electron-hole pair). At wavelengths blueward of the quantum yield threshold, it is possible for multiple carriers to be produced, and then (by diffusion) end up in separate wells, leading to an additional contribution to the flat field autocorrelation function \citep[e.g.,][]{2008PASP..120..759M} as well as errors in gain determination.

\subsection{Perfect detector}
\label{ss:D1}

In a perfect detector, with $\alpha$, $\beta$, and ${a}$ all zero, each pixel operates independently. The mean charge accumulated in pixel $(i,j)$ in frame $a$ is
\begin{equation}
\langle Q_a(i,j) \rangle = It_a
\label{eq:Poisson1}
\end{equation}
and the covariance structure is
\begin{equation}
{\rm Cov}\left[ Q_a(i,j), Q_b(i+\Delta i,j+\Delta j) \right] = I t_{\min(a,b)} \delta_{\Delta i,0} \delta_{\Delta j,0}.
\label{eq:Poisson2}
\end{equation}
Since the signal difference $S_{ab}(i,j) = g^{-1}[Q_b(i,j)-Q_a(i,j)]$, we then find a covariance structure:
\begin{equation}
C_{abcd}(\Delta i,\Delta j) = \frac{I}{g^2} \left[ t_{\min(a,c)} - t_{\min(a,d)} - t_{\min(b,c)} + t_{\min(b,d)} \right] \delta_{\Delta i,0} \delta_{\Delta j,0}.
\label{eq:Cabcd-ideal}
\end{equation}
It is convenient then to define
\begin{equation}
t_{abcd} \equiv t_{\min(a,c)} - t_{\min(a,d)} - t_{\min(b,c)} + t_{\min(b,d)}
~~~{\rm and}~~~
t_{ab} \equiv t_b-t_a;
\label{eq:tabcd-def}
\end{equation}
by inspection if $a< b$ and $c< d$, then $t_{abcd}$ is the amount of time in the intersection of the intervals $(t_a,t_b)\cap(t_c,t_d)$. In this case, we also have $t_{abab}=t_{ab}$.

Equation~(\ref{eq:Cabcd-ideal}) is behind the usual concept of obtaining a system gain from a variance vs.\ mean plot: we have
\begin{equation}
C_{abab}(0,0) = \frac{I}{g^2} t_{ab} ~~~{\rm and}~~~ \langle S_{ab}(i,j) \rangle = \frac Igt_{ab}.
\end{equation}
For a detector with no read noise, the ratio of variance to mean is then $1/g$. In practice $C_{abab}(0,0)$ contains a contribution from read noise, which can be removed by taking the slope of the variance vs.\ mean plot. 

We now consider the various non-ideal detector effects. We begin by considering the effects one at a time, but we also need to consider interactions between the IPC and non-linearity, i.e.\ effects of order $\alpha\beta$ and $\alpha a$.

\subsection{Inter-pixel capacitance}
\label{ss:D2}

In the presence of IPC, the covariance structure of Eq.~(\ref{eq:Cabcd-ideal}) is modified via smoothing by the IPC kernel. The IPC kernel acts locally in time, so we may write
\begin{eqnarray}
C_{abcd}(\Delta i,\Delta j) &=& \langle S_{ab}(i,j) S_{cd}(i+\Delta i,j+\Delta j) \rangle
\nonumber \\
&=& \sum_{ i_1, j_1, i_2,  j_2} K_{i_1,j_1} K_{i_2,j_2} \langle S_{ab}(i-i_1,j-j_1) S_{cd}(i-i_2+\Delta i,j-j_2+\Delta j) \rangle
\nonumber \\
&=& \sum_{ i_1, j_1, i_2,  j_2} K_{i_1,j_1} K_{i_2,j_2} \frac I{g^2} t_{abcd}
\delta_{(i-i_1)-(i-i_2+\Delta i),0}\delta_{(j-j_1)-(j-j_2+\Delta j),0}
\nonumber \\
&=& \sum_{ i_1, j_1} K_{i_1,j_1} K_{i_1+\Delta i,j_1+\Delta j} \frac I{g^2} t_{abcd}.
\label{eq:IPC-corr-general}
\end{eqnarray}
If the IPC kernel is represented by nearest-neighbor parameters $\alpha_{\rm H,V}$, and diagonal-neighbor parameters $\alpha_{\rm D}$, then
\begin{eqnarray}
C_{abcd}(0,0) &=& \frac I{g^2} t_{abcd} [ (1-4\alpha - 4\alpha_{\rm D})^2 + 2\alpha_{\rm H}^2+2\alpha_{\rm V}^2 + 4\alpha_{\rm D}^2],
\nonumber \\
C_{abcd}(\pm 1,0) &=& \frac I{g^2} t_{abcd} \left[ 2\alpha_{\rm H}(1-4\alpha-4\alpha_{\rm D}) + 4\alpha_{\rm V}\alpha_{\rm D} \right],
\nonumber \\
C_{abcd}(0,\pm 1) &=& \frac I{g^2} t_{abcd} \left[ 2\alpha_{\rm V}(1-4\alpha - 4\alpha_{\rm D}) + 4\alpha_{\rm H}\alpha_{\rm D} \right],
{\rm ~~and}
\nonumber \\
C_{abcd}(\pm 1,\pm 1) = C_{abcd}(\pm 1,\mp 1) &=& \frac I{g^2} t_{abcd} \left[ 2\alpha_{\rm H}\alpha_{\rm V} + 2\alpha_{\rm D}(1-4\alpha-4\alpha_{\rm D}) \right].
\label{eq:IPC-corr}
\end{eqnarray}
(There are other non-zero terms.) The nearest-neighbor correlations are thus useful for measuring $\alpha_{\rm H}$ and $\alpha_{\rm V}$, and the diagonal-neighbor correlations for $\alpha_{\rm D}$.

Note that regardless of $K$, the IPC-induced correction to $C_{abcd}(\Delta i,\Delta j)$ remains proportional to $t_{abcd}$. Therefore, if IPC is the only non-ideal effect in the detector, the correlation function will be zero if $t_{abcd}=0$. The ``disjoint correlation functions'' with $t_{abcd}=0$ are therefore diagnostics of other effects -- including, as we shall see, the brighter-fatter effect.

\subsection{Classical non-linearity}
\label{ss:D3}

The classical non-linearity -- that arising from the nonlinearity of the electrons to data numbers conversion, Eq.~(\ref{eq:S-g-beta}) -- contributes a correction to the correlation function that involves the third moment of the Poisson distribution. In the presence of only classical non-linearity, but no IPC or BFE, the pixels still operate independently, so for simplicity we will consider only one pixel. The connected skewness\footnote{Connected skewnesses are defined by $\langle ABC\rangle_{\rm conn} = \langle \Delta A\Delta B\Delta C\rangle$, where $\Delta A \equiv A-\langle A\rangle$, etc.} of charges at different times is
\begin{equation}
\langle Q_a(i,j) Q_b(i,j) Q_c(i,j) \rangle_{\rm conn} = It_{\min(a,b,c)},
\end{equation}
since the connected skewness of a Poisson distribution is its mean, and all counts received after $t_{\min(a,b,c)}$ are independent of $Q_{\min(a,b,c)}$. This leads, after some algebra, to the ancillary result
\begin{equation}
{\rm Cov} [Q_a(i,j),\, Q_b(i,j)^2] = 2I^2t_b t_{\min(a,b)} + It_{\rm min(a,b)}.
\label{eq:Poisson3}
\end{equation}

Then we find (suppressing $i$ and $j$ indices to avoid clutter):
\begin{eqnarray}
C_{abcd}(0,0) &=& \frac1{g^2} {\rm Cov} \left\{
Q_b - \beta Q_b^2 -Q_a + \beta Q_a^2,\,
Q_d - \beta Q_d^2 -Q_c + \beta Q_c^2
\right\}
\nonumber \\
&=& \frac1{g^2} \left\{
{\rm Cov}(Q_a,Q_c) - {\rm Cov}(Q_a,Q_d) - {\rm Cov}(Q_b,Q_c) + {\rm Cov}(Q_b,Q_d)\right\}
\nonumber \\ &&
+ \frac\beta{g^2}\Bigl\{
{\rm Cov}(Q_a^2,Q_d) - {\rm Cov}(Q_a^2,Q_c)
-{\rm Cov}(Q_b^2,Q_d) + {\rm Cov}(Q_b^2,Q_c)
\nonumber \\ &&~~
+{\rm Cov}(Q_c^2,Q_b) - {\rm Cov}(Q_c^2,Q_a)
-{\rm Cov}(Q_d^2,Q_b) + {\rm Cov}(Q_d^2,Q_a)
\Bigr\}
\nonumber \\
&=& \frac1{g^2} It_{abcd} - 2\frac{\beta}{g^2} It_{abcd}
\nonumber \\
&& + 2\frac{\beta I^2}{g^2}
\Bigl\{
(t_a+t_d) t_{\min(d,a)}
-(t_b+t_d) t_{\min(d,b)}
-(t_a+t_c) t_{\min(c,a)}
+(t_b+t_c) t_{\min(c,b)}
\Bigr\}.
\nonumber \\
&&
\end{eqnarray}
We will define
\begin{equation}
T_{abcd} \equiv
-(t_a+t_d) t_{\min(d,a)}
+(t_b+t_d) t_{\min(d,b)}
+(t_a+t_c) t_{\min(c,a)}
-(t_b+t_c) t_{\min(c,b)}
\label{eq:_Tabcd-def}
\end{equation}
(units: s$^2$) so that
\begin{equation}
C_{abcd}(0,0)  = \frac1{g^2}(1-2\beta) It_{abcd} - 2\frac\beta{g^2}I^2T_{abcd}.
\label{eq:beta-cont}
\end{equation}
Here the ``$1-2\beta$'' correction term is of little interest, since the correction is tiny even compared to WFIRST requirements -- indeed, it represents the nonlinearity generated by a single electron, and if $\beta\sim {\cal O}(1)$ ppm/e, then this is a correction of order $10^{-6}$. The $T_{abcd}$ term can be much larger.

Note the following special cases of $T_{abcd}$:
\begin{list}{$\bullet$}{}
\item
If $a\le b\le c\le d$, then we have $t_{abcd}=0$ and $T_{abcd} = t_{ab}t_{cd}\ge 0$.
\item
If $a\le c\le b\le d$, then we have $t_{abcd}=t_{bc}$ and $T_{abcd} = t_{ab}t_{cd} + (t_b+t_c)t_{bc}\ge 0$.
\item
If $a=c$ and $b=d$, then we have $t_{abcd}=t_{ab}$ and $T_{abcd}=2t_bt_{ab}$.
\end{list}

\subsection{Interdependence of IPC and non-linearity}
\label{ss:D4}

Because the IPC corrections to flat results are often large (e.g.\ $\alpha=1.25\%$ leads to an $\sim 10$\%\ correction to the gain!) we need to consider the way in which IPC interacts with the non-linearity curve. This is particularly true given that IPC-non-linearity interactions affect both of the flat auto-correlation measurements of the BFE presented in this document. In particular, we want to capture the order $\alpha\beta$ terms in the flat auto-correlation function.

\changetext{IPC and non-linearity may interact in a complicated way because the two steps do not in general commute. The approach taken here is the mathematical point of view: one chooses an ordering, and any additional effects -- including issues associated with order of operations -- are packaged into ``non-linear IPC'' (\S\ref{ss:D6}). The ordering we choose here is IPC first and then non-linearity (consistent with ``standard'' pipelines that treat non-linearity as the last step in the signal chain and thus the first correction implemented in data processing). From the physical point of view, the non-linear capacitance coming from the depletion region in the photodiode and the capacitive links between pixels should be thought of as a non-linear capacitor network that is solved simultaneously (the non-linearities coming from the rest of the signal chain and the analog-to-digital converter really do come later). Further consideration of this physical point of view is deferred to future work.}

\changetext{With these assumptions, the} non-linearity interacts with the IPC according to
\begin{eqnarray}
[S_{\rm initial} - S_{\rm final}](i,j) \!\!&=&\!\! \frac1g\Bigl\{
\sum_{\Delta i,\Delta j} K_{\Delta i,\Delta j} Q(i+\Delta i,j+\Delta j)
- \beta \Bigl[
\sum_{\Delta i,\Delta j} K_{\Delta i,\Delta j} Q(i+\Delta i,j+\Delta j)
\Bigr]^2 \Bigr\}
\nonumber \\
&\approx &\!\! \frac1g\bigl\{
(1-4\alpha-4\alpha_{\rm D}) Q(i,j) + \alpha_{\rm H}[Q(i+1,j) + Q(i-1,j)]
\nonumber \\ &&
 + \alpha_{\rm V}[Q(i,j+1) + Q(i,j-1)]
\nonumber \\ &&
 + \alpha_{\rm D}[Q(i+1,j+1) + Q(i+1,j-1) + Q(i-1,j+1) + Q(i-1,j-1)]
\nonumber \\ &&
- \beta (1- 8\alpha) Q^2(i,j)
 - 2\alpha_{\rm H}\beta Q(i,j) [Q(i+1,j) + Q(i-1,j)]
\nonumber \\ &&
- 2\alpha_{\rm V}\beta Q(i,j) [Q(i,j+1) + Q(i,j-1)]
\bigr\},
\label{eq:Sij-ab}
\end{eqnarray}
where the approximation includes terms of order $\alpha_{\rm H,V}\beta$ but not $\alpha_{\rm D}\beta$ or $\alpha^2\beta$.

Our principal interest is in the contributions of order $\alpha\beta$ to the correlation function, which occur at either zero lag $(\Delta i,\Delta j)=(0,0)$ or for nearest-neighbor pixels, $(\Delta i,\Delta j)\in{\cal N}$. In general, the contribution of order $\alpha\beta$ to $C_{abcd}(\Delta i,\Delta j)$ (denoted below as $\Delta C_{abcd}(\Delta i,\Delta j)|_{\alpha\beta}$)
has four parts: the covariance of the order $\alpha\beta$ term in $(i,j)$ with the order $1$ term in $(i+\Delta i,j+\Delta j)$ (which we will call the ``$\alpha\beta\times 1$'' term); the $\alpha\times\beta$ term; the $\beta\times\alpha$ term; and the $1\times\alpha\beta$ term. These can be read off from Eq.~(\ref{eq:Sij-ab}), and covariances can be computed using the fact that (i) the {\em charges} in each pixel are independent, and (ii) the Poisson statistics needed are in Eqs.~(\ref{eq:Poisson1}), (\ref{eq:Poisson2}), and (\ref{eq:Poisson3}); this is an algebraically lengthy but straightforward exercise. The result is
\begin{equation}
\Delta C_{abcd}(0,0)|_{\alpha\beta} = \frac{\alpha\beta}{g^2} ( 16 I^2 T_{abcd} + 12 I t_{abcd})
\label{eq:bias-00}
\end{equation}
for zero lag,
\begin{equation}
\Delta C_{abcd}(\pm 1,0)|_{\alpha\beta} = -\frac{4\alpha_{\rm H}\beta}{g^2} (I^2 T_{abcd} + It_{abcd})
\label{eq:bias-H}
\end{equation}
for the horizontal neighbors, and
\begin{equation}
\Delta C_{abcd}(0,\pm 1)|_{\alpha\beta} = -\frac{4\alpha_{\rm V}\beta}{g^2} (I^2 T_{abcd} + It_{abcd}).
\label{eq:bias-V}
\end{equation}
for the vertical nearest neighbors. (The order $\alpha_{\rm H}\beta$ and $\alpha_{\rm V}\beta$ contributions beyond the 4 nearest neighbor pixels are zero.)
Note that we normally have $I^2T_{abcd}\gg It_{abcd}$, so that term is dominant.

\subsection{Brighter-fatter effect; moving pixel boundaries}
\label{ss:D5}

\changetext{The effect of the BFE on pixel correlation functions in infrared arrays can be treated by considering the charge $Q(i,j;t)$ at time $t$ in pixel $(i,j)$ as a stochastic function. The new ingredient is that the charge deposited in each pixel between $t$ and $t+\delta t$ depends on the charge already present at time $t$. The BFE process is Markovian, in the sense that the state at $t+\delta t$ depends on the state at time $t$, but has no further dependence on the state of the system at any earlier times.\footnote{\changetext{If we also tried to include charge trapping effects such as persistence, then they would not be Markovian. These are not treated in the present formalism.}} This allows us to write the moments of the charge (here we consider the first two moments, the mean and covariance) at time $t+\delta t$ in terms of those at time $t$; by working to order $\delta t$, we can construct a system of differential equations for the moments, which we solve starting from the initial condition at $t=0$. This is analogous to what has been done for CCDs (see \citealt{2018AJ....155..258C} for an implementation of a very similar method; see also \citealt{2019arXiv190508677A}, who work directly with derivatives rather than very short but finite time steps). We extend this method to apply to unequal-time correlation functions by writing a similar set of differential equations for the covariance between a pixel at time $t_1$ and another pixel at time $t\ge t_1$, again with $t$ as the independent variable. We can take the covariance at $t_1$ as an initial condition.}

Let us define the area defect of a pixel at time $t$ to be
\begin{equation}
W(i,j;t) \equiv 1 + \sum_{\Delta i,\Delta j} {a}_{\Delta i,\Delta j} Q({i+\Delta i,j+\Delta j},t);
\label{eq:W}
\end{equation}
this is close to 1, with deviations controlled by the BFE.
Then -- given the state of the system $Q(i,j;t)$ at time $t$ -- we can find the mean charge in pixel $(i,j)$ at time $t+\dt$ as
\begin{equation}
\left\langle Q(i,j;t + \dt) \right\rangle|_{t} = Q(i,j;t) + I W(i,j;t) \dt,
\label{eq:Q1}
\end{equation}
where $\dt$ is taken to be small, and the subscript $|_t$ denotes that the state of the detector at time $t$ is fixed. Here $ I W(i,j;t) \dt$ is the probability that an electron is collected in pixel $(i,j)$ between times $t$ and $t+\dt$; we assume $I\dt\ll 1$ (one electron at a time), and will take the limit as $\dt\rightarrow 0$ so that this approximation becomes arbitrarily good. \changetext{Since we are turning the result into a first-order differential equation for each moment, we can drop terms of order $\delta t^2$ and higher in what follows.} The change in 2nd moment is
\begin{eqnarray}
\left\langle Q(i,j;t + \dt)Q(i',j';t + \dt) \right\rangle|_{t}
&=& Q(i,j;t)Q(i',j';t) + I W(i,j;t) Q(i',j';t) \dt
\nonumber \\
&&+ I W(i',j';t) Q(i,j;t) \,\dt + I W(i,j;t) \delta_{ii'}\delta_{jj'} \,\dt,
\nonumber \\
&&
\label{eq:Q2}
\end{eqnarray}
where we have expanded $Q(i,j;t + \Delta t) = Q(i,j;t ) +\Delta Q(i,j;t)$, and the four terms on the right hand side correspond to the expectation values of $Q(i,j;t)Q(i',j';t)$, $\Delta Q(i,j;t)Q(i',j';t)$, $Q(i,j;t)\Delta Q(i',j';t)$, and $\Delta Q(i,j;t) \Delta Q(i',j';t)$ respectively. The last term is only non-zero if the two pixels are identical ($\delta_{ii'}\delta_{jj'}=1$), since then a single electron can increment both $Q(i,j)$ and $Q(i',j')$.

It is now possible to solve the above system of equations to first order in ${a}$. Let us first consider Eq.~(\ref{eq:Q1}). Taking the average of the right-hand side over possible realizations at time $t$, we see that
\begin{equation}
\left\langle Q(i,j;t + \dt) \right\rangle= \langle Q(i,j;t)\rangle + I\, \dt + \sum_{\Delta i,\Delta j} {a}_{\Delta i,\Delta j} \langle Q({i+\Delta i,j+\Delta j},t)\rangle \,I\, \dt.
\end{equation}
Recalling that $\suma = \sum_{\Delta i,\Delta j} {a}_{\Delta i,\Delta j}$, and using translation invariance to show that the $Q(i,j;t)$ all have the same expectation value, we see that
\begin{equation}
\left\langle Q(i,j;t + \dt) \right\rangle= \langle Q(i,j;t)\rangle + I \,\dt + \suma \langle Q(i,j;t)\rangle I\, \dt.
\end{equation}
This becomes a differential equation for $\langle Q(i,j;t)\rangle$:
\begin{equation}
\frac{d}{dt} \langle Q(i,j;t)\rangle = I (1 + \suma \langle Q(i,j;t)\rangle),
\end{equation}
with solution starting from $\langle Q(i,j;t)\rangle=0$ at $t=0$:
\begin{equation}
\langle Q(i,j;t)\rangle = \frac{e^{I\suma t}-1}{\suma} \approx It + \frac12\suma I^2 t^2.
\label{eq:Q1-sol}
\end{equation}
(The approximation holds to first order in the ${a}$ coefficients.)

The next step is to solve for the covariance matrix. This proceeds in two steps. First, one tracks the full second moment from time 0 to some later time $t_1$. Then one tracks a conditional second moment to a later time $t_2\ge t_1$. The mean of Eq.~(\ref{eq:Q2}) is
\begin{eqnarray}
\left\langle Q(i,j;t + \Delta t)Q(i',j';t + \Delta t) \right\rangle
&=& \left\langle Q(i,j;t)Q(i',j';t)\right\rangle + I \left\langle Q(i',j';t)\right\rangle \Delta t
+ I \left\langle Q(i,j;t)\right\rangle \Delta t
\nonumber \\
&& + I \delta_{ii'}\delta_{jj'} \Delta t
+ I \sum_{\Delta i,\Delta j} {a}_{\Delta i,\Delta j} \Bigl[
\left\langle Q(i+\Delta i,j+\Delta j;t) Q(i',j';t)\right\rangle
\nonumber \\
&&
+\left\langle Q(i'+\Delta i,j'+\Delta j;t) Q(i,j;t)\right\rangle
\nonumber \\
&&
+\delta_{ii'}\delta_{jj'} \left\langle Q(i+\Delta i,j+\Delta j;t)\right\rangle
\Bigr] \Delta t.
\end{eqnarray}
This can be turned into a differential equation. The first moment solution from Eq.~(\ref{eq:Q1-sol}) can be substituted in, and all second order terms in ${a}$ dropped:
\begin{eqnarray}
\frac{d}{dt} \left\langle Q(i,j;t)Q(i',j';t) \right\rangle
&=&  
2I^2t + \suma I^3 t^2
+ I \delta_{ii'}\delta_{jj'} + I^2\suma t \delta_{ii'}\delta_{jj'}
\nonumber \\
&& 
+ I \sum_{\Delta i,\Delta j} {a}_{\Delta i,\Delta j} \Bigl[
\left\langle Q(i+\Delta i,j+\Delta j;t) Q(i',j';t)\right\rangle
\nonumber \\
&&
+\left\langle Q(i'+\Delta i,j'+\Delta j;t) Q(i,j;t)\right\rangle
\Bigr].
\label{eq:temp.1}
\end{eqnarray}
The initial condition is that $\left\langle Q(i,j;t)Q(i',j';t) \right\rangle=0$ at $t=0$.
To solve Eq.~(\ref{eq:temp.1}) to first order in $a$, we use the standard method of first solving the equation at $a=0$ (the zeroth order solution), then substituting this into any term multiplying $a$ (or $\suma$) and solving again. This gives
\begin{equation}
\left\langle Q(i,j;t)Q(i',j';t) \right\rangle
=  
I^2t^2 + \suma I^3 t^3
+ It \delta_{ii'}\delta_{jj'} + \frac12 I^2\suma t^2 \delta_{ii'}\delta_{jj'}
+ \frac12({a}_{i-i',j-j'} + {a}_{i'-i,j'-j})I^2t^2.
\label{eq:Q2-sol.1}
\end{equation}

In our case, however, we need not just equal-time but also unequal-time correlation functions of the charge. This means we need to propagate the second moment at time $t_1$ to the \changetext{covariance between times $t_1$ and $t_2$ (without loss of generality, $t_2>t_1$). Since the system is Markovian, if $t_1\le t$ we can take the expectation value in Eq.~(\ref{eq:Q1}) to be conditioned not only on the state of the system at $t$ but also at $t_1$. Then we can multiply Eq.~(\ref{eq:Q1}) by $Q(i',j';t_1)$. Since $Q(i',j';t_1)$ is fully determined by the state of the system at $t_1$, we can pull it inside the expectation value:}
\begin{equation}
\changetext{\left\langle Q(i',j';t_1) Q(i,j;t + \dt) \right\rangle|_{t_1,t} = Q(i',j';t_1) Q(i,j;t) + I W(i,j;t) Q(i',j';t_1) \dt.}
\label{eq:Q1-condition}
\end{equation}
\changetext{We next average this over states of the system at $t_1$ and $t$ (i.e., we remove the condition, but then get an expectation value on the right-hand side).}
Turning \changetext{the result} into a differential equation, we have
\begin{eqnarray}
\frac{d}{dt} \left\langle Q(i,j;t)Q(i',j';t_1)\right\rangle
&=& I \left\langle W(i,j;t)Q(i',j';t_1) \right\rangle
\nonumber \\
&=& I \left\langle Q(i',j';t_1) \right\rangle + I \sum_{\Delta i,\Delta j} {a}_{\Delta i,\Delta j}
\left\langle Q(i+\Delta i,j+\Delta j;t)Q(i',j';t_1) \right\rangle.
\nonumber \\ &&
\label{eq:difeq-temp.1}
\end{eqnarray}
with initial condition from Eq.~(\ref{eq:Q2-sol.1}), $\left\langle Q(i,j;t_1)Q(i',j';t_1) \right\rangle|_{{a}=0} = I^2t_1^2 + It_1 \delta_{ii'}\delta_{jj'}$.
The solution to first order in $a$ is
\begin{eqnarray}
\left\langle Q(i,j;t)Q(i',j';t_1) \right\rangle
&=&
I^2t_1t
+ \frac12\suma I^3t_1t(t+t_1)
+ \left( It_1 + \frac12 I^2\suma t_1^2\right) \delta_{ii'}\delta_{jj'}
\nonumber \\ &&
+ \frac12({a}_{i-i',j-j'} + {a}_{i'-i,j'-j})I^2t_1^2
+ {a}_{i'-i,j'-j} I^2t_1(t-t_1).
\label{eq:Q2-sol}
\end{eqnarray}
Subtracting out $\left\langle Q(i,j;t)\right\rangle\left\langle Q(i',j';t_1) \right\rangle$ from Eq.~(\ref{eq:Q1-sol}) gives the covariance matrix:
\begin{eqnarray}
{\rm Cov}\left[ Q(i,j;t), Q(i',j';t_1) \right]
&=&
\left( It_1 + \frac12 I^2\suma t_1^2\right) \delta_{ii'}\delta_{jj'}
\nonumber \\ &&
+ \frac12({a}_{i-i',j-j'} + {a}_{i'-i,j'-j})I^2t_1^2
+ {a}_{i'-i,j'-j} I^2t_1(t-t_1).
\end{eqnarray}
Recall that this is for $t\ge t_1$; for $t<t_1$, one can use the symmetry of the covariance matrix to obtain the result.
If one considers only the linear response of the detector, this maps directly into the flat autocorrelation function:
\begin{eqnarray}
C_{abcd}(\Delta i,\Delta j) \!\!\!\!
&=& \!\!\frac{1}{g^2} \Bigl\{
{\rm Cov}\left[ Q(i,j;t_a), Q(i+\Delta i,j + \Delta j;t_c) \right]
-{\rm Cov}\left[ Q(i,j;t_a), Q(i+\Delta i,j + \Delta j;t_d) \right]
\nonumber \\ &&
-{\rm Cov}\left[ Q(i,j;t_b), Q(i+\Delta i,j + \Delta j;t_c) \right]
+{\rm Cov}\left[ Q(i,j;t_b), Q(i+\Delta i,j + \Delta j;t_d) \right]
\Bigr\}
\nonumber \\
&=& \!\!\frac{1}{g^2} \Bigl\{
\Bigl[ It_{abcd} + \frac12 I^2\suma \sigma_{abcd}\Bigr] \delta_{\Delta i,0}\delta_{\Delta j,0}
+ \frac12({a}_{\Delta i,\Delta j} + {a}_{-\Delta i,-\Delta j})I^2 t_{ab} t_{cd}
\nonumber \\ &&
- \frac12({a}_{\Delta i,\Delta j} - {a}_{-\Delta i,-\Delta j})I^2\tau_{abcd}
 \Bigr\},
\label{eq:Cabcd-gamma}
\end{eqnarray}
where we define the auxiliary quantities
\begin{equation}
\sigma_{abcd} = t^2_{\min(a,c)} - t^2_{\min(a,d)} - t^2_{\min(b,c)} + t^2_{\min(b,d)}
\label{eq:sigma-abcd-def}
\end{equation}
and
\begin{equation}
\tau_{abcd} = t_{ac}t_{\min(a,c)} - t_{ad}t_{\min(a,d)} - t_{bc}t_{\min(b,c)} + t_{bd}t_{\min(b,d)}.
\label{eq:tau-abcd-def}
\end{equation}
Here $\sigma_{abcd}$ and $\tau_{abcd}$ have units of s$^2$ and satisfy the following rules:
\begin{list}{$\bullet$}{}
\item
$\sigma_{abcd} = \sigma_{cdab}$
and $\tau_{abcd} = -\tau_{cdab}$.
\item
If $a\le b\le c\le d$, then $\sigma_{abcd} = 0$ and $\tau_{abcd} = t_{ab}t_{cd}\ge 0$.
\item
If $a=c \le b=d$, then $\sigma_{abcd} = t_{ab}(t_a+t_b)$ and $\tau_{abcd} = 0$.
\end{list}
Note that in Eq.~(\ref{eq:Cabcd-gamma}), $\tau_{abcd}$ describes the response of a correlation function to the odd part of ${a}$ while the response to the even part is described by $t_{ab} t_{cd}$. The response to the summed effect $\suma$ is encoded in $\sigma_{abcd}$.

In the presence of IPC, the BFE contribution to the correlation function should be convolved twice with the IPC kernel:
\begin{equation}
C^{\rm BFE\,with\,IPC}_{abcd}(\Delta i,\Delta j) = 
\sum_{i_1,j_1,i_2,j_2} K_{i_1,j_1} K_{i_2,j_2}
C^{\rm BFE\,without\,IPC}_{abcd}(\Delta i+i_1+i_2, \Delta j+j_1+j_2).
\end{equation}

\subsection{Non-linear inter-pixel capacitance (NL-IPC)}
\label{ss:D6}

The contribution of NL-IPC to the covariance of signals is, to order $K'$,
\begin{eqnarray}
&& \!\!\!\!\!\!\!\!\!\!\!\!\!\!\!\!
{\rm Cov}[S_a(i,j), S_c(i+\Delta i,j+\Delta j)]|_{K'}
\nonumber \\
&=& \frac1{g^2} \Bigl\{
\sum_{i',j'}
K'_{i-i',j-j'} K_{i+\Delta i-i', j+\Delta j-j'}
\bar Q_a {\rm Cov}[Q_a(i',j'), Q_c(i',j')]
\nonumber \\ &&
\sum_{i',j'}
K'_{i+\Delta i-i',j+\Delta j-j'} K_{i-i', j-j'}
\bar Q_c {\rm Cov}[Q_a(i',j'), Q_c(i',j')]
\Bigr\}
\nonumber \\
&=&
\frac1{g^2} [KK']_{\Delta i,\Delta j} I^2 (t_a+t_c) t_{\min(a,c)},
\end{eqnarray}
where in the first expression the first term comes from the order $K'$ contribution to $S_a(i,j)$ and the second term from the contribution to $S_c(i+\Delta i,j+\Delta j)$. The final expression used the symmetry of $K'_{\Delta i,\Delta j}$ under $(\Delta i,\Delta j)\rightarrow(-\Delta i,-\Delta j)$, and has defined the convolution $[KK']_{\Delta i,\Delta j} = \sum_{i_1,j_1} K_{i_1,j_1} K'_{\Delta i-i_1,\Delta j-j_1}$. The contribution to the correlation function is
\begin{equation}
C_{abcd}(\Delta i,\Delta j)|_{K'} = \frac1{g^2} [KK']_{\Delta i,\Delta j} I^2 T_{abcd}.
\label{Cabcd-NLIPC}
\end{equation}

\subsection{Combined correlation function}
\label{ss:combined-cf}

Putting together all of the combinations -- IPC, non-linearity, BFE, NL-IPC, and the leading order interactions -- we have the following expression, including corrections of order $\alpha$, $\alpha^2$, $\beta$, $\alpha\beta$, $a$, $\alpha a$, $\alpha'$, and $\alpha\alpha'$:
\begin{eqnarray}
C_{abcd}(\Delta i,\Delta j) \!\!\!\!
&=& \!\!\frac{1}{g^2} \Bigl\{
\Bigl( It_{abcd} + \frac12 I^2\suma \sigma_{abcd}\Bigr) [K^2]_{\Delta i,\Delta j}
+ \frac12([K^2a]_{\Delta i,\Delta j} + [K^2a]_{-\Delta i,-\Delta j})I^2 t_{ab} t_{cd}
\nonumber \\ &&
- \frac12([K^2a]_{\Delta i,\Delta j} - [K^2a]_{-\Delta i,-\Delta j})I^2\tau_{abcd}
- 2\beta (I^2T_{abcd} + It_{abcd} ) \delta_{\Delta i,0}\delta_{\Delta j,0}
\nonumber \\ &&
+ \alpha\beta (16 I^2 T_{abcd} + 12 It_{abcd})\delta_{\Delta i,0}\delta_{\Delta j,0}
- 4\alpha_{\rm H}\beta (I^2T_{abcd} + It_{abcd} ) \delta_{|\Delta i|,1}\delta_{\Delta j,0}
\nonumber \\ &&
- 4\alpha_{\rm V}\beta (I^2T_{abcd} + It_{abcd} ) \delta_{\Delta i,0}\delta_{|\Delta j|,1}
+ [KK']_{\Delta i,\Delta j} I^2 T_{abcd}
 \Bigr\},
\label{eq:master-cf}
\end{eqnarray}
where we have defined $[K^2]_{\Delta i,\Delta j}$ to be the auto-convolution of $K$, and $[K^2a]_{\Delta i,\Delta j}$ to be the convolution of $K^2$ and $a$.

\subsection{Special cases used in detector characterization}
\label{ss:specialcases}

We now turn our focus to the special cases that are used in detector characterization. We consider two special cases of the correlation function: the equal-interval correlation functions ($a=c<b=d$ -- most similar to the auto-correlation that one would obtain from a CCD) and the non-overlapping correlation functions ($a<b<c<d$ -- which exhibits new features only accessible with a non-destructive read capability). We also consider the mean-variance plot, which is a common diagnostic of the gain of a detector system, with a particular emphasis on how IPC, nonlinearity, and BFE affect the gain measurement.

In what follows, terms of order $\alpha$, $\alpha^2$, $\beta$, ${a}$, $\alpha\beta$, and $\alpha a$ are kept. Higher terms in the non-ideal detector effects are dropped.

\subsubsection{Equal-interval correlation function}

The case of $a=c<b=d$ corresponds to the auto-correlation of a single difference image $S_a-S_b$. It is therefore most comparable to what one would obtain with a CCD. The contributions at zero lag sum to:
\begin{eqnarray}
C_{abab}(0,0) &=& \frac I{g^2} t_{ab} \Bigl\{ (1-4\alpha-4\alpha_{\rm D})^2 + 2(\alpha_{\rm H}^2+\alpha_{\rm V}^2) + 4\alpha_{\rm D}^2
- 4(1-8\alpha)\beta I t_b
- 2(1-6\alpha)\beta
\nonumber \\ &&
+ [K^2a]_{0,0} It_{ab}
+ \frac12 (1-8\alpha) \suma I (t_a+t_b)
+ 2[KK']_{0,0} I t_b
 \Bigr\} ,
\label{eq:auto-zero}
\end{eqnarray}
while the horizontal nearest neighbors are
\begin{eqnarray}
C_{abab}(\pm 1,0) &=& \frac I{g^2} t_{ab} \Bigl\{ 2\alpha_{\rm H}(1-4\alpha-4\alpha_{\rm D}) + 4\alpha_{\rm V}\alpha_{\rm D}
 -8\alpha_{\rm H}\beta \left( It_b + \frac12 \right)
 + \alpha_{\rm H} \suma I(t_a+t_b)
\nonumber \\ &&
  + [K^2a]_{\rm H} It_{ab}
 +2 [KK']_{1,0} I t_b
   \Bigr\},
\label{eq:auto-nearest}
\end{eqnarray}
where we define ${a}_{\rm H} = ({a}_{1,0}+{a}_{-1,0})/2$. A similar equation holds for the vertical nearest neighbors. For the diagonal neighbors, we have
\begin{equation}
C_{abab}(\langle 1, 1\rangle) = \frac I{g^2} t_{ab} \Bigl\{ 2\alpha_{\rm D}(1-4\alpha-4\alpha_{\rm D}) + 4\alpha_{\rm H}\alpha_{\rm V}
  + [K^2a]_{\langle 1,1\rangle} It_{ab}
 +2 [KK']_{\langle 1,1\rangle} I t_b
   \Bigr\}.
\label{eq:auto-diagonal}
\end{equation}

The equal-interval correlation function, especially but not exclusively at zero lag, contains a large contribution from read noise (from various sources), and this must be removed before interpreting it.

The time-translation-averaged versions of Eqs.~(\ref{eq:auto-zero}--\ref{eq:auto-diagonal}) can be evaluated with straightforward algebra; they are
\begin{eqnarray}
\bar C_{abab[n]}(0,0) \!\!\!\! &=&\!\!\!\! \frac I{g^2} t_{ab} \Bigl\{ (1-4\alpha-4\alpha_{\rm D})^2 + 2(\alpha_{\rm H}^2+\alpha_{\rm V}^2) + 4\alpha_{\rm D}^2
- 4(1-8\alpha)\beta I\left( t_b + \frac{n-1}2\Delta t \right)
\nonumber \\ &&\!\!\!\!
- 2(1-6\alpha)\beta
+ [K^2a]_{0,0} It_{ab}
+ \frac12 (1-8\alpha) \suma I [t_a+t_b + (n-1)\Delta t]
\nonumber \\ &&\!\!\!\!
+ 2[KK']_{0,0} I \left( t_b + \frac{n-1}2\Delta t \right)
 \Bigr\}
\nonumber \\ &&
\label{eq:auto-zero-shift}
\end{eqnarray}
for zero lag;
\begin{eqnarray}
\bar C_{abab[n]}(\pm 1,0) \!\!\!\!&=&\!\!\!\! \frac I{g^2} t_{ab} \Bigl\{ 2\alpha_{\rm H}(1-4\alpha-4\alpha_{\rm D}) + 4\alpha_{\rm V}\alpha_{\rm D}
 -8\alpha_{\rm H}\beta \left( It_b + \frac{n-1}2 I\Delta t + \frac12 \right)
\nonumber \\ &&\!\!\!\!
 + \alpha_{\rm H} \suma I[t_a+t_b+(n-1)\Delta t]
  + [K^2a]_{\rm H} It_{ab}
 +2 [KK']_{1,0} I \left( t_b + \frac{n-1}2\Delta t \right)
   \Bigr\}~~~~
\label{eq:auto-nearest-shift}
\end{eqnarray}
for the nearest neighbor; and
\begin{eqnarray}
\bar C_{abab[n]}(\langle 1,1\rangle) \!\!\!\!&=&\!\!\!\! \frac I{g^2} t_{ab} \Bigl\{ 2\alpha_{\rm D}(1-4\alpha-4\alpha_{\rm D})
  + [K^2a]_{\langle 1,1\rangle} It_{ab}
 +2 [KK']_{\langle 1,1\rangle} I \left( t_b + \frac{n-1}2\Delta t \right)
   \Bigr\}~~~~
\label{eq:auto-diagonal-shift}
\end{eqnarray}
for the diagonal neighbor.

\subsubsection{Non-overlapping correlation function}

The case of $a<b<c<d$ is special, because then $\sigma_{abcd}=t_{abcd}=0$ and the correlation function -- including contributions of IPC, classical non-linearity, and BFE -- simplifies to
\begin{eqnarray}
C_{abcd}(\Delta i,\Delta j)|_{a<b<c<d}
\!\! &=& \!\! \frac{I^2 t_{ab}t_{cd}}{g^2} \Bigl\{ [K^2a]_{-\Delta i,-\Delta j}
+ [KK']_{\Delta i,\Delta j}
- 2(1-8\alpha)\beta\delta_{\Delta i,0}\delta_{\Delta j,0}
\nonumber \\ &&
- 4 \alpha_{\rm H}\beta\delta_{|\Delta i|,1}\delta_{\Delta j,0}
- 4 \alpha_{\rm V}\beta\delta_{\Delta i,0}\delta_{|\Delta j|,1}
\Bigr\}.
\label{eq:uet-master}
\end{eqnarray}
That is, the non-overlapping correlation function is directly sensitive to the coefficients ${a}_{\Delta i,\Delta j}$, has no sensitivity to linear IPC at order $\alpha$, and only has sensitivity to the classical non-linearity $\beta$ at zero lag. At order $\alpha\beta$, there is a contribution in the nearest neighbors. There is a trivial mapping from the pixel-space lag in the correlation function $(\Delta i,\Delta j)$ to the lag in the BFE kernel ${a}_{\Delta i,\Delta j}$. Thus this should be a ``clean'' measurement of the inter-pixel non-linear effects (BFE and NL-IPC), insensitive to small errors in the determination of $I$ and $g$. Any source of noise that is uncorrelated across frames is also removed. The reset ($kTC$) noise is also removed, since the correlation function is constructed from correlated double samples. The main drawback is that the method is only sensitive to a combination of BFE and NL-IPC, and cannot distinguish between the two mechanisms.

The one large correction that is necessary is that ${a}_{0,0}$ must be corrected for the classical non-linearity (which is a larger effect than the BFE). Therefore we need to measure $\beta$ from the non-linearity ($t^2$ term) of the signal vs.\ time plot of the flat. Interestingly, this is sensitive to the combination $\beta-\frac12\suma$. It follows that the flat non-linearity and the non-overlapping correlation functions contain an intrinsic degeneracy where $\beta$ and ${a}_{0,0}$ are both changed but holding the combination $\beta-\frac12{a}_{0,0}$ constant. Other correlation functions are needed to break this degeneracy.

A secondary correction is that $C_{abcd}(\Delta i,\Delta j)$ in Eq.~(\ref{eq:uet-master}) is the correlation function of the signal, but the measurement contains signal+noise. Therefore any noise that is correlated across frames must be characterized and removed from $C_{abcd}(\Delta i,\Delta j)$.

\subsubsection{Mean-variance slope}

A common method to estimate the gain of a system is to determine the ratio of the mean signal in a pair of matched flats to the variance. In practice, since the measured variance contains read noise, one measures the slope of the variance as a function of the mean, e.g.:
\begin{equation}
\hat g^{\rm raw}_{abcd} \equiv \frac{M_{cd}-M_{ab}}{V_{cd}-V_{ab}},
\label{eq:g-raw}
\end{equation}
where $M_{ab} = \langle S_a(i,j)-S_b(i,j)\rangle$ and $V_{ab} = C_{abab}(0,0)$ is the variance of a difference frame. This construction only makes sense for $(a,b)\neq(c,d)$ (a common case is $a=c<b<d$). 
The mean is
\begin{equation}
M_{ab} = \frac{It_{ab}}g \left[ 1 - \left(\beta - \frac\suma2\right) I (t_a+t_b) \right].
\label{eq:Mab-NL}
\end{equation}
We obtain the variance from Eq.~(\ref{eq:auto-zero}). The mean-variance slope is related to the gain by
\begin{eqnarray}
\hat g^{\rm raw}_{abcd}
\!\! &=& \!\! \frac{g}{(1-4\alpha-4\alpha_{\rm D})^2 + 2(\alpha_{\rm H}^2+\alpha_{\rm V}^2) + 4 \alpha_{\rm D}^2}\Bigl\{
1
+ 2 \beta I \frac{ t_{cd}t_d  - t_{ab} t_b }{t_{cd}-t_{ab}}
\nonumber \\ &&
+ \left[\beta+(1+8\alpha)[K^2a]_{0,0}\right] I(t_{cd}+t_{ab})
+ 2(1+2\alpha)\beta
\nonumber \\ &&
+ 2(1+8\alpha)[KK']_{0,0} I \frac{t_{cd}t_d - t_{ab}t_b}{t_{cd}-t_{ab}}
\Bigr\}.
\label{eq:MV-slope-abcd}
\end{eqnarray}
In the special case of $a=c<b<d$ (which will be used herein), we find
\begin{eqnarray}
\hat g^{\rm raw}_{abad}
\!\! &=& \!\! \frac{g}{(1-4\alpha-4\alpha_{\rm D})^2 + 2(\alpha_{\rm H}^2+\alpha_{\rm V}^2) + 4 \alpha_{\rm D}^2}\Bigl\{
1
+ \left[ 2 \beta - 8(1+3\alpha)\alpha' \right] I t_a
\nonumber \\ &&
+ \left[ 3 \beta -(1+8\alpha)[K^2a]_{0,0} + 8(1+3\alpha)\alpha' \right] I(t_{ad}+t_{ab})
+ 2(1+2\alpha)\beta
\Bigr\}.
\label{eq:MV-slope-abd}
\end{eqnarray}
Here, we have used that to order $\alpha\alpha'$,
\begin{equation}
(1+8\alpha) [KK']_{0,0}
\approx -4(1+4\alpha)\alpha' + 4\alpha\alpha' = -4(1+3\alpha)\alpha'.
\end{equation}

In Eq.~(\ref{eq:MV-slope-abd}), the pre-factor is the traditional IPC correction to the gain. Following this is a non-linear correction term that depends on the ``start time'' $t_a$ of the measurement. Then comes a second non-linear correction term that depends on the ``duration'' $t_{ad}+t_{ab}$ of the measurement. Both are proportional to $\beta$ (or to the ${a}_{\Delta i,\Delta j}$); they have the same dependence in the special case of $t_a=t_c=0$. The last term is formally of order $\beta$, but is smaller than the previous two correction terms as it does not contain a factor of accumulated charge $It$.

\section{Flat field simulations}
\label{sec:sim}

\begin{figure}[h]
\centering
\includegraphics[height=8.0in]{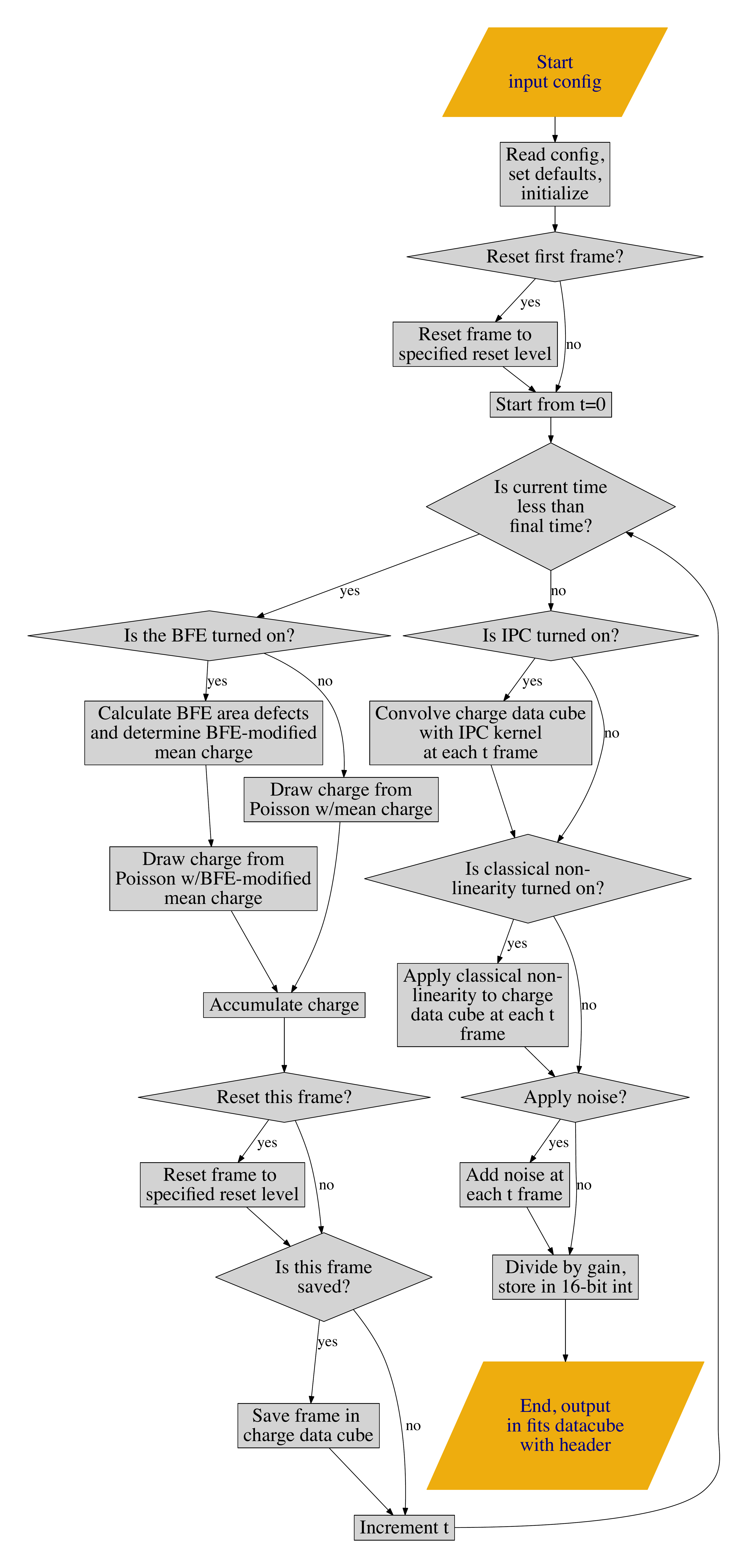}
\caption{\label{fig:flow} Flowchart showing the construction procedure for basic flat simulations containing the BFE, IPC, and classical-nonlinearity.}
\end{figure}

We construct simulations for validation and interpretation.  This simulated data set contains flats and darks that are designed to resemble the real data cubes from the Detector Characterization Laboratory (DCL) at NASA Goddard Space Flight Center, with an implementation of the key effects described in the earlier sections of this paper.  The procedure consists of three main steps, visualized in a flowchart in Fig.~\ref{fig:flow}: first, user inputs and specifications are read from a configuration file (top); second, charge is accumulated via draws from a Poisson distribution and modified by a BFE kernel, if the BFE is turned on (loop on lower left); third, all other effects including linear IPC, classical non-linearity, and noise are applied to each time step of the charge array, which is ultimately converted to a signal and stored in an output fits data cube (lower right). The remainder of this section delves into the specifics of how the simulated flat fields are constructed.

\subsection{Details of simulation procedure}

The first part of the script sets up the simulation that will be created by ingesting a configuration file and using defaults when selections are not specified by the user.  The default settings create a datacube with dimensions of $4096^2$ pixels$^{2}$ with 66 time samples with the bounding 4 rows and columns designated as reference pixels. Substeps set the total number of time slices at which the charge is computed between the stored time slices with the default set to \texttt{substep=2} (for this default the computation is done for $2 \times 66 = 112$ time steps). This setting exists to ensure convergence when the BFE mode is turned on.  The user specifies quantities like gain $g$, current per pixel per second $I$, length of time sample in seconds, quantum efficiency QE, and IPC $\alpha$.  Reset frames and reset levels can also be set in the config file.

After reading in user specifications and initializing arrays, charge is drawn and accumulated over the total time steps.  For the initial time frame, a random realization of charge is drawn from a Poisson distribution with a mean of $QE \times I \times \delta t$, with $\delta t$ being the time between each time step.  If the BFE is turned on, a matrix of pixel area defects $W(i,j;t)$ given by Eq.~\ref{eq:W} is calculated by convolving a user-specified input kernel $\bfea$ with the charge distribution over the pixel grid at the given time $t$.  Subsequent time frames compound the previous time frames with charge drawn from a Poisson distribution with the mean modified by the pixel area defects, i.e. charge is drawn from a Poisson distribution with a mean of $W \times QE \times I \times \delta t$.  If the BFE is turned off, all time frames are accumulated with fully random realizations of charge.

After the charge has accumulated over all time frames, a linear IPC can be applied by convolving the full charge data cube with an IPC kernel.  Note that from this stage onward, operations are only performed on the time samples that will be saved (i.e. in the default settings, IPC is applied to the 66 time samples and not the intermediate substeps).  Non-linearity $\beta$ can also be applied after the IPC, where $Q(i,j;t) \rightarrow Q(i,j;t)-\beta [Q(i,j;t)]^2$.

We create realizations of noise datacubes using \textsc{nghxrg}\footnote{https://github.com/BJRauscher/nghxrg}, the HxRG Noise Generator written in Python by Bernard Rauscher \citep{2015PASP..127.1144R}.  This software produces white read noise, pedestal drifts, correlated and uncorrelated pink noise, alternating column noise, and picture frame noise and was based on a principle components analysis of the James Webb Space Telescope NIRSpec detector subsystem. Here, we use input parameters tuned to WFIRST configurations.

The final step is to convert the charge into DN by dividing by $g$ and save in an array of unsigned 16-bit integers.  The output datacube is saved in fits format with a header containing information about the configuration settings and parameter values used to run the simulation.

The main flat field generation is done as part of the \textsc{solid-waffle} pipeline, which is further described in Section~\ref{sec:char}; the noise file must be generated separately using the \textsc{nghxrg} package.

\subsection{Test bed of simulated flats and darks}

We created a set of 10 simulated flat fields and 10 simulated darks to test the characterization and BFE measurement framework presented in this paper.  The input parameters were chosen to resemble the real detector data analyzed in Paper II and are summarized in the `truth' column of Table~\ref{tab:sim_summary}.  All simulated data cubes are ascending ramps where the signal level in DN increases as time increases and have \texttt{NAXIS1=NAXIS2=4096} (spatial dimensions) and \texttt{NAXIS3=66} (readout time frames).  The time between each time frame is 2.75 s, $g=2.06$, QE=0.95, and $\beta=0.58$.  The IPC kernel is as described in Eq.~\ref{eq:norm-K} so that $K_{0,0} = 1-4\alpha$, $K_{0,\pm 1} = K_{\pm 1,0} = \alpha$, and all others are zero, with $\alpha=0.0169$. The BFE is turned on, and has a zero-lag component $a_{0,0}=-1.372$ ppm/e. We conservatively set \texttt{substep=20} to ensure convergence. Table~\ref{tab:sim_summary} provides symmetrized mean values of $[K^2a]_{\Delta i \Delta j}$, the convolution of $K^2$ (the auto-convolution of K) and $\bfea$.  The simulated flats each have illumination $I=559$ e/s/pixel, while the simulated darks have illumination $I=0.191$ e/s/pixel, which was chosen so that the resulting slope of the signal vs time matched a typical real dark.  Random seeds from \texttt{1001-1010} and \texttt{2001-2010} were set for the flats and darks, respectively.

Finally, we generated 10 noise data cubes using \textsc{nghxrg} with \texttt{NAXIS} dimensions as specified above, \texttt{n\_out=32} (number of detector outputs), \texttt{nroh=8} (row overhead in pixels), \texttt{rd\_noise=4} (standard deviation of white read noise in e), \texttt{pedestal=4} (pedestal drift in e), \texttt{c\_pink=3} (standard deviation of correlated pink noise), \texttt{u\_pink=1} (standard deviation of uncorrelated pink noise), \texttt{c\_ACN=1} (standard deviation of alternating column noise).  For simplicity, we did not add any picture frame noise, and we set a bias offset of 19222 e to match a typical real dark.  See \citet{2015PASP..127.1144R} for further details of the noise recipe. Each of the 10 noise data cubes are combined with a flat and dark each so that each flat/dark pair have a noise realization in common; this simplification should not affect the results of this analysis in any significant way.  These simulations were run on Pitzer \citep{Pitzer2018}, a supercomputer at the Ohio Supercomputing Center \citep{OhioSupercomputerCenter1987}.

\section{Characterization based on flat fields}
\label{sec:char}

We now turn to the practical problem at hand: extracting the calibration parameters ($g$, $\alpha$, $\beta$, $a_{\Delta i,\Delta j}$, etc.) from a suite of flat field and dark exposures. We first provide an overview of our characterization pipeline (known as \textsc{solid-waffle}), and then describe in detail the modules therein. The tools are written in Python 2, with data stored in \textsc{numpy} arrays. Due to the large file size associated with flat fields using multiple up-the-ramp samples (2.2 GB {\em per file} for a 66-frame H4RG flat), the full data set is not stored in RAM; instead the \textsc{fitsio} package was used to enable rapid access to small subsets of the data from disk without reading the entire file.\footnote{With standard FITS routines and ``usual astronomer writing Python'' level of attention to data handling, reading the files can completely swamp the computation time!}

Our analysis takes as input $N$ flat fields and $N$ dark images, where $N\ge 2$. The SCA is broken into a grid of $N_x\times N_y$ ``super-pixels,'' each of size $\Delta_x \times \Delta_y$ physical pixels. Statistical properties such as medians, variances, and correlation functions are understood to be computed in each super-pixel. Note that $N_x\Delta_x = N_y\Delta_y = 4096$ for an H4RG (and 2048 for an H2RG). Super-pixels may be made larger to improve S/N, but this implies more averaging over the SCA so localized features and patterns may be washed out (we will see examples of this in Paper II).

Each super-pixel is processed through ``basic'' characterization. Following this, it passes through inter-pixel non-linearity (IPNL) determination using the non-overlapping correlation function, and then (optionally) through advanced characterization and other tests. We now describe these steps.

\subsection{Basic characterization}
\label{ss:char-basic}

The basic characterization step for a super-pixel is a prerequisite to studying all of the more subtle effects in the NIR detectors. It uses four time frames $t_a$, $t_b$, $t_c$, and $t_d$, and it does not take into account the diagonal IPC, the brighter-fatter effect, non-linear IPC, or signal-dependent QE.

We first construct the CDS images $S_{ab}(i,j|{\rm F}_k)$ and $S_{ad}(i,j|{\rm F}_k)$ within the range of column $i$ and row $j$ in the super-pixel, for each flat F$_k$. We build a median (over flats $k$) image $f(i,j) = {\rm med}_{k=1}^N S_{ad}(i,j|{\rm F}_k)$, and then a pixel mask based on requiring $f(i,j)$ to be within 10\% of its median (this time taken over $i,j$). This rejects disconnected or low-response pixels.

Our next step is to perform a reference pixel subtraction. The procedure used here was obtained after some experimentation with DCL data, and is the default in our code, but may require some re-adjustment for other setups. We first find the range of rows $j_{\rm min} ... j_{\rm max}=j_{\rm min}+\Delta_y-1$ corresponding to the super-pixel, and find the two $4\times \Delta_y$ blocks of reference pixels on the left and right sides of the SCA. For each flat exposure ${\rm F}_k$, and for each of our two CDS difference images ($S_{ab}$ and $S_{ad}$), we find the median of these $8\Delta_y$ pixels, and subtract this from the entire super-pixel. A similar procedure is applied to the dark images D$_k$. Note that this procedure only adds or subtracts a constant in the super-pixel, and does {\em not} correct each individual row.\footnote{Correcting each row would print noise from the reference pixels as additional horizontal correlations. There are row-dependent drifts in the electronics, however we found that these are better eliminated at the correlation function level by either subtracting the correlation function in the darks or by the ``baseline subtraction'' method described in \S\ref{ss:char-ipnl}.}

We next want to compute the raw gain, $\hat g_{abad}^{\rm raw}$. To do this, we need to compute the mean signal levels $M_{ab}$ and $M_{ad}$. The current default is to take the reference-corrected image $S_{ab}(i,j|{\rm F}_k)$, and compute a mean in $k$ followed by a median in $(i,j)$. The variance $V_{ab}$ is obtained by taking each of the $N(N-1)/2$ pairs of flats $(k,\ell)$, with $1\le k<\ell\le N$. For each pair, we compute the difference $S_{ab}(i,j|{\rm F}_k) - S_{ab}(i,j|{\rm F}_\ell)$, and compute the inter-quartile range (IQR) of the $\Delta_x\Delta_y$ pixels.\footnote{We use difference images because they are robust against permanent structure in the flat fields, e.g.\ variations in pixel area or quantum efficiency.} The variance is estimated as $({\rm IQR}/1.349)^2/2$, as appropriate for a Gaussian (but note that the IQR estimator is robust against outliers, unlike the standard variance estimator), and with a factor of 2 to account for the fact that the flat difference has noise from both flats. The $V_{ab}$ used in the raw gain estimator is the average of the $N(N-1)/2$ estimates obtained from the various flat pairs. These means and variances are then plugged into Eq.~(\ref{eq:g-raw}).

Inter-pixel capacitance is addressed through the flat field auto-correlation method, which we implement as follows. For each of the $N(N-1)/2$ flat pairs, we construct a difference $T(i,j|{\rm F}_k,{\rm F}_\ell) = S_{ad}(i,j|{\rm F}_k) - S_{ad}(i,j|{\rm F}_\ell)$. We clip the top $100\epsilon$\% and bottom $100\epsilon$\% of the $T(i,j|{\rm F}_k,{\rm F}_\ell)$ map, leaving $100(1-2\epsilon)$\% of the pixels unmasked. Then we define a horizontal correlation
\begin{equation}
C_{\rm H}(|{\rm F}_k,{\rm F}_\ell) = \frac1{\#\,{\rm pix}\,(i,j)} \sum_{(i,j)} \left\{
[T(i,j|{\rm F}_k,{\rm F}_\ell)-\bar T(|{\rm F}_k,{\rm F}_\ell)]
[T(i+1,j|{\rm F}_k,{\rm F}_\ell)-\bar T(|{\rm F}_k,{\rm F}_\ell)]
\right\},
\end{equation}
where the average is over pixels where both that pixel $(i,j)$ and its horizontal neighbor $(i+1,j)$ are unmasked. We then compute an averaged horizontal correlation
\begin{equation}
C_{\rm H} = \frac12 \times \frac1{f_{\rm corr}} \times \frac1{N(N-1)/2}\sum_{1\le k<\ell\le N}\left[
C_{\rm H}(|{\rm F}_k,{\rm F}_\ell) - C_{\rm H}(|{\rm D}_k,{\rm D}_\ell)\right].
\end{equation}
Here we have subtracted the correlation from a pair of dark frames (to remove the contribution of correlated read noise), and averaged over the flat pairs. The factor of $\frac12$ takes into account the fact that by subtracting two flats, we have doubled the correlation function. Finally, the factor of $f_{\rm corr}$ takes into account the suppression of correlations by the histogram clipping of $T$. It depends on $\epsilon$; for our default choice of $\epsilon = 0.01$, we have $f_{\rm corr} = 0.7629$. See Appendix~\ref{app:covar-clip} for a derivation of $f_{\rm corr}$. A similar calculation is used to obtain the vertical correlation function $C_{\rm V}$ and the diagonal correlation function $C_{\rm D}$.

Finally, we need a measure of ramp curvature. We construct the difference box
\begin{equation}
R(i,j|{\rm F}_k) = S_{cd}(i,j|{\rm F}_k) - \frac{t_{cd}}{t_{ab}} S_{ab}(i,j|{\rm F}_k)
\end{equation}
and perform the usual reference pixel subtraction (based on all $8\Delta_y$ ``left+right'' reference pixels in the same range of rows as the super-pixel). We clip the pixels corresponding to the top and bottom $100\epsilon$\% of the histogram of $S_{ab}(i,j|{\rm F}_k)$ and of $R(i,j|{\rm F}_k)$, and then compute
\begin{equation}
{\tt frac\_dslope} = \frac{\sum_{k=1}^N R(i,j|{\rm F}_k) / t_{cd}}{\sum_{k=1}^N S_{ab}(i,j|{\rm F}_k) / t_{ab}}.
\end{equation}
Note that $1+{\tt frac\_dslope}$ is the ratio of the slope of the signal (in DN/frame) in the $cd$ interval relative to the $ab$ interval. For a perfectly linear detector, ${\tt frac\_dslope}$ should be zero. For a non-linear detector, the mean signal is
$\langle S_a(i,j)\rangle = [It_a - \beta_{\rm r}(It_a)^2]/g$,
where
\begin{equation}
\beta_{\rm r} = \beta - \frac12\suma
\label{eq:beta-r}
\end{equation}
is the ramp curvature (here $\suma$ denotes the signal-dependent QE, and enters via Eq.~\ref{eq:Q1-sol}). Via straightforward algebra, we can see that the slope difference ratio ${\tt frac\_dslope}$ is expected to be $- \beta I (t_c+t_d-t_a-t_b)$.

We may now construct an IPC + non-linearity corrected ($\alpha\beta$-corrected) gain $g$, estimated current per pixel $I$, horizontal and vertical IPC $\alpha_{\rm H}$ and $\alpha_{\rm V}$, and ramp curvature $\beta_{\rm r}$ by iteratively solving the system of equations:
\begin{eqnarray}
\hat g^{\rm raw}_{abad} &=& g \frac{ 1 + \beta_{\rm r} I (3t_b+3t_d-4t_a) }
{ (1-2\alpha_{\rm H}-2\alpha_{\rm V})^2 + 2\alpha_{\rm H}^2 + 2\alpha_{\rm V}^2 };
\nonumber \\
C_{\rm H} &=& \frac{2 I t_{ad} \alpha_{\rm H} }{g^2} (1 - 2\alpha_{\rm H} - 2\alpha_{\rm V} - 4 \beta_{\rm r} I t_d );
\nonumber \\
C_{\rm V} &=& \frac{2 I t_{ad} \alpha_{\rm V} }{g^2} (1 - 2\alpha_{\rm H} - 2\alpha_{\rm V} - 4 \beta_{\rm r} I t_d );
\nonumber \\
M_{ad} &=& \frac{ It_{\rm ad} }{g} [ 1 - \beta_{\rm r} I (t_a+t_d) ];
~~{\rm and}
\nonumber \\
{\tt frac\_dslope} &=& - \beta_{\rm r} I (t_c+t_d-t_a-t_b).
\end{eqnarray}
This is 5 equations for 5 unknowns; note that the difference between $\beta$ and $\beta_{\rm r}$ (i.e., the signal-dependent QE term $\suma$) has been neglected in the gain and IPC determination. Initializing the system with $g=\hat g^{\rm raw}_{abad}$, $\alpha_{\rm H} = \alpha_{\rm V} = \beta = 0$, $I=gM_{ad}/t_{ad}$, and solving the above equations in turn for $g$, $\alpha_{\rm H}$, $\alpha_{\rm V}$, $I$, and $\beta$ leads to rapid convergence.

The resulting parameters $g$, $\alpha_{\rm H}$, $\alpha_{\rm V}$, $I$, and $\beta$ contain small residual biases due to the BFE, nonlinear IPC, and signal-dependent QE if these phenomena are present. These will be explored in more detail in Paper II.

\subsection{IPNL determination via the non-overlapping correlation function}
\label{ss:char-ipnl}

With the basic parameters in each super-pixel measured, we may now measure the non-overlapping correlation function, $C_{abcd}(\Delta i,\Delta j)$ for $a<b<c<d$. This is almost a direct test for the presence of inter-pixel non-linearities (BFE and NL-IPC), since it contains no contribution from linear IPC, and only small corrections for classical non-linearity ($\beta$) are required. In particular, at zero lag, Eq.~(\ref{eq:uet-master}) can be rearranged to give
\begin{equation}
[K^2a]_{0,0} + [KK']_{0,0} = \frac{g^2}{I^2 t_{ab} t_{cd}}C_{abcd}(0,0) + 2(1-8\alpha)\beta.
\end{equation}
The ramp curvature does not yield an estimate directly for $\beta$, but rather the combination $\beta_{\rm ramp} = \beta-\frac12\suma$. We also recall that to order $\alpha a$, we have $[K^2a']_{0,0} = [K^2a]_{0,0} - (1-8\alpha)\suma$. We can thus write, to ${\cal O}({\alpha a})$:
\begin{equation}
[K^2a']_{0,0} + [KK']_{0,0} = \frac{g^2}{I^2 t_{ab} t_{cd}}C_{abcd}(0,0) + 2(1-8\alpha)\beta_{\rm r}.
\label{eq:M1-g00}
\end{equation}
Similarly, one may compute the adjacent pixel correlation functions:
\begin{equation}
[K^2a']_{\pm 1,0} + [KK']_{\pm1,0} = \frac{g^2}{I^2 t_{ab} t_{cd}}C_{abcd}(\mp 1,0) + 4\alpha_{\rm H}\beta_{\rm r},
\label{eq:M1-gH}
\end{equation}
and similarly for the vertical directions.

Equations~(\ref{eq:M1-g00}) and (\ref{eq:M1-gH}) show that the non-overlapping correlation function method, as we have implemented it, is sensitive to the $[K^2a']_{\Delta i,\Delta j}+ [KK']_{\Delta i,\Delta j}$ coefficients. Note that the BFE and NL-IPC appear together, both with $t_{ab} t_{cd}$ time dependence, and the non-overlapping correlation function method provides no way to separate them. This method has only a small correction on the right-hand side due to the ramp curvature $\beta_{\rm r}$, so this method of IPNL determination is not subject to spurious detection due to small errors in the basic parameters ($g$, $I$, and $\alpha$). In most practical situations, we will find that the $4\alpha_{\rm H}\beta_{\rm r}$ correction is smaller than the IPNL, and the $2(1-8\alpha)\beta_{\rm r}$ correction is similar to the IPNL (see Paper II for quantitative details on a WFIRST development detector).

Our pipeline provides results out to a separation of 2 pixels in either the horizontal or vertical directions, i.e., it reports a $5\times 5$ kernel $[K^2a'+KK']$.

\begin{figure}[t!]
\includegraphics[width=6.2in]{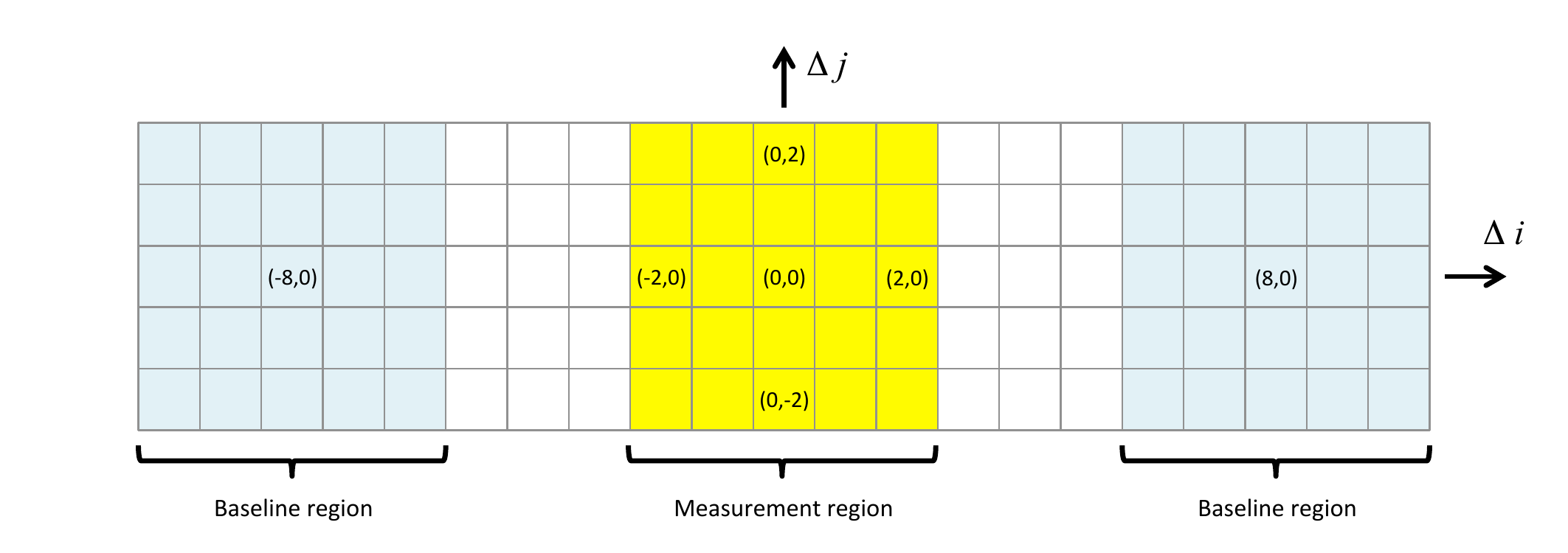}
\caption{\label{fig:bc}The baseline correction scheme used in Eq.~(\ref{eq:BC}). We carry out measurements of the non-overlapping correlation function $C_{abcd}(\Delta i,\Delta j)$ with pixels at separation $(\Delta i,\Delta j)$. We are interested in measurements of the BFE in a $5\times 5$ pixel region centered on zero lag (yellow shaded region). The ``baseline'' is measured in the blue shaded regions; each yellow measurement pixel is corrected using blue baseline pixels in the same row. The fast-read direction is horizontal.}
\end{figure}

We now turn to the implementation details of $C_{abcd}(\Delta i,\Delta j)$ in the pipeline itself. The correlation function can be determined by the same methods used to compute $C_{\rm H}$ and $C_{\rm V}$. However, we found in initial studies on DCL data that the measurements showed statistically significant deviations depending on which flat was used, which are suspected to be low frequency noise in the data (see horizontal stripes in the dark image and discussion in Paper II). Therefore, the default setting in our pipeline is to filter out the low frequencies via a baseline correction: instead of using the raw correlation function,
\begin{equation}
C^{\rm raw}_{abcd}(\Delta i,\Delta j) = \frac1{N_{\rm pair}} \sum_{i,j} \left\{ [S_a-S_b](i,j) - \overline{S_a-S_b} \right\}
\left\{ [S_c-S_d](i+\Delta i,j+\Delta j) - \overline{S_c-S_d} \right\}
\end{equation}
(where the overbar denotes an average and $N_{\rm pair}$ is the number of pixel pairs in the sum), we find a ``baseline'' contribution:
\begin{equation}
C^{\rm baseline}_{abcd}(\Delta j) = \frac1{N'_{\rm pair}} \sum_{i,j,\Delta i'} \left\{ [S_a-S_b](i,j) - \overline{S_a-S_b} \right\}
\left\{ [S_c-S_d](i+\Delta i',j+\Delta j) - \overline{S_c-S_d} \right\},
\label{eq:BL}
\end{equation}
where the pair summation runs over $6\le |\Delta i'|\le 10$, and again $N'_{\rm pair}$ is the number of pixel pairs in the sum. That is, the baseline is the correlation function obtained by replacing pixel $(i+\Delta i,j+\Delta j)$ with the average of pixels in the same row but 6--10 pixels left or right (ahead or behind in the readout sequence). Both the leading and trailing regions are used with equal weight, {\em except} that (i) the standard 1\%\ outlier rejection is used before taking the covariance, and (ii) the implementation in the code rejects one of these regions if the pixel pair $(i,j)\leftrightarrow (i+\Delta i',j+\Delta j)$ would span an output channel boundary. The correction regions are shown schematically in Figure~\ref{fig:bc}.
We then define a corrected correlation function:
\begin{equation}
C^{\rm corrected}_{abcd}(\Delta i,\Delta j)
= C^{\rm raw}_{abcd}(\Delta i,\Delta j) - C^{\rm baseline}_{abcd}(\Delta i, \Delta j).
\label{eq:BC}
\end{equation}

\subsection{Advanced characterization}
\label{ss:char-adv}

While the basic characterization stage is sufficient to provide a pixel mask and the properties (gain, IPC, non-linearity) needed to convert the non-overlapping correlation function to an IPNL measurement, there are several ways it could be improved. The statistical uncertainties in the gain and IPC are significant, especially with small super-pixels. Moreover, if the BFE exists in these detectors (and we will see in Paper II that it does), then it imprints a bias on $g$, $\alpha$, etc., and an iterative process is required to de-bias the final result. The ``advanced characterization'' tool handles both of these issues.

To motivate our approach to the first issue (noise in the parameters), and understand the improvement in knowledge of gain and IPC that can be achieved, let us first recall the uncertainty in gain and IPC achievable by the ``basic'' approach. If $t_{ab}\ll t_{ad}$, then in the computation of $\hat g_{abad}^{\rm raw}$ from Eq.~(\ref{eq:g-raw}), the uncertainty is dominated by $V_{ad}$. The variance of a Gaussian distribution with $n_{\rm pix} = \Delta_x\Delta_y$ samples has a fractional uncertainty of $\sqrt{2/n_{\rm pix}}$. Similarly, the correlation coefficient $\rho\sim 2\alpha$ of two adjacent pixels has $2n_{\rm pix}$ samples (counting both vertical and horizontal pairs) and hence an uncertainty of $\sqrt{1/(2n_{\rm pix})}$. With $N-1$ flat pairs, we should thus in principle achieve
\begin{equation}
\left.\frac{\sigma(g)}{g}\right|_{\rm perfect} \approx \sqrt{\frac{2}{(N-1)n_{\rm pix}}}
~~{\rm and}~~
\left.\sigma(\alpha)\right|_{\rm perfect} \approx \sqrt{\frac{1}{2(N-1)n_{\rm pix}}}.
\label{eq:err-ideal}
\end{equation}
In practice, our pipeline does not do this quite well -- the uncertainty in $V_{ab}$ is not negligible, and the use of the IQR carries a factor of 1.64 penalty in error for a Gaussian relative to the ``idealized'' case.\footnote{See, e.g., \cite{book.DasGupta2011}, \S9.5 for a general discussion of this issue.} However, if the flat field has $N_{\rm frame}$ samples, and we break it into ``sub-flats'' of length $\mu$, one might expect that by combining the sub-flats we could achieve an uncertainty that is reduced by a factor of $\sqrt{N_{\rm frame}/\mu}$. Our pipeline does not quite achieve this, but it nevertheless can beat the estimate in Eq.~(\ref{eq:err-ideal}). One expects that if $\mu$ is decreased, we should see a reduction in the error (down to the fundamental limit of $\mu=1$). However, the magnitudes of the correlation functions decrease as one decreases $\mu$, and hence we become more sensitive to the subtraction of noise from $C_{\rm H}$ and $C_{\rm V}$. Therefore there is a trade-off in the choice of $\mu$ (and $\mu'$ defined below), and we allow the user to set these in the configuration file.

The implementation of these ideas in our pipeline is as follows. First, the user sets the range of frames used (earliest frame $a$ and latest frame $d$), as well as two integers $\mu$ and $\mu'$ (with $\mu'<\mu$) in the configuration file describing the spacing of time slices used in the gain and IPC determination; typical values would be $\mu'=1$ and $\mu=3$. We then compute an averaged correlation function
\begin{equation}
\bar C_{\rm H} = \bar C_{a,a+\mu,a,a+\mu,[d-a-\mu]}(\pm 1,0) = \frac1{d-a-\mu+1} \sum_{j=0}^{d-a-\mu} C_{{\rm H}, a+j, a+j+\mu}, 
\end{equation}
where $C_{{\rm H}, a+j, a+j+\mu}$ is obtained using the same methodology as in basic characterization using the difference image of frames $a+j$ and $a+j+\mu$. Something similar is performed to compute $\bar C_{\rm V}$. Finally, we compute the difference of variances
\begin{equation}
\Delta\bar V = \bar C_{a,a+\mu,a,a+\mu,[d-a-\mu]}(0,0)
- \bar C_{a,a+\mu',a,a+\mu',[d-a-\mu]}(0,0)
= \frac{ \sum_{j=0}^{d-a-\mu} (V_{a+j,a+j+\mu} - V_{a+j,a+j+\mu'})}{d-a-\mu+1},
\end{equation}
where $V_{ef}$ is the variance of the difference of frames $e$ and $f$ as obtained using the same methodology as in basic characterization.

In the advanced characterization stage, the mean information on the ramp is obtained by taking the sequence of differences $M_{a,a+1}$, $M_{a+1,a+2}$, ... $M_{d-1,d}$, and performing a linear fit:
\begin{equation}
M_{j,j+1} =  c_0 + c_1j + {\rm residuals},
\end{equation}
where the sum of the square of residuals is minimized. One then wants to simultaneously solve the equations:
\begin{eqnarray}
\Delta\bar V &=& \frac{I \Delta t}{g^2} [(1-4\alpha)^2 + 2\alpha_{\rm H}^2 + 2\alpha_{\rm V}^2] (\mu-\mu') - 4(1-8\alpha)\beta_{\rm r} \frac{(I\Delta t)^2}{g^2}  \bigl[ \mu(a+\mu) - \mu'(a+\mu')
\nonumber \\ && + \frac{d-a-\mu}{2}(\mu-\mu') \bigr] + {\rm Err}[\Delta\bar V],
\nonumber \\
\bar C_{\rm H} &=& 2 \frac{I \Delta t }{ g^2}\mu \left[ 1-4\alpha - 4\alpha_{\rm D} - 4 \beta_{\rm r} \left( \frac{d+a+\mu}{2} I\Delta t + \frac12 \right) \right] \alpha_{\rm H} + 4\frac{I \Delta t }{ g^2}\mu \alpha_{\rm V}\alpha_{\rm D} + {\rm Err}[\bar C_{\rm H}],
\nonumber \\
\bar C_{\rm V} &=& 2 \frac{I \Delta t }{ g^2}\mu \left[ 1-4\alpha - 4\alpha_{\rm D} - 4 \beta_{\rm r} \left( \frac{d+a+\mu}{2} I\Delta t + \frac12 \right) \right] \alpha_{\rm V} + 4\frac{I \Delta t }{ g^2}\mu \alpha_{\rm H}\alpha_{\rm D} + {\rm Err}[\bar C_{\rm V}],
\nonumber \\
\bar C_{\rm D} &=& 2 \frac{I \Delta t }{ g^2}\mu \left[ (1-4\alpha-4\alpha_{\rm D})\alpha_{\rm D} \right] \alpha_{\rm V}
+ 2 \frac{I \Delta t }{ g^2}\mu \alpha_{\rm H}\alpha_{\rm V} + {\rm Err}[\bar C_{\rm D}],
\nonumber \\
c_1 &=& -2 \beta_{\rm r} \frac{(I\Delta t)^2}{g} + {\rm Err}[c_1],
~~~{\rm and}
\nonumber \\
c_0 &=& \frac{1}{g}[I\Delta t-\beta_{\rm r}(I\Delta t)^2] + {\rm Err}[c_0].
\end{eqnarray}
Here ``Err[...]'' denotes the contribution to the specified quantity coming from BFE, NL-IPC, and signal-dependent QE (we will consider these shortly; in future versions of the pipeline we may add other effects). Once again, these are 6 equations for 6 unknowns ($g$, $I$, $\alpha_{\rm H}$, $\alpha_{\rm V}$, $\alpha_{\rm D}$, and $\beta_{\rm r}$). A straightforward and effective method is to alternately use the $\Delta\bar V$, $c_1$, and $c_0$ equations to solve algebraically for $I$, $g$, and $\beta_{\rm r}$; and then to use the $\bar C_{\rm H}$, $\bar C_{\rm V}$, and $\bar C_{\rm D}$ equations to solve for $\alpha_{\rm H}$, $\alpha_{\rm V}$, and $\alpha_{\rm D}$.

The advanced characterization pipeline can run in two modes for computing the error terms Err[...]; these are {\tt none}, {\tt bfe}, and {\tt nlipc}. The {\tt none} mode is the simplest: it sets the error terms to zero. When run on a detector that has, e.g., the BFE, the {\tt none} mode is subject to similar biases as the ``basic'' characterization, but can give smaller statistical error.

Given that we will see in Paper II that the BFE is significant for the H4RGs, we included the {\tt bfe} mode. This computes the error terms Err[...] under the assumption that there is a BFE ($a_{\Delta i,\Delta j}\neq 0$), but with no non-linear IPC ($K'_{\Delta i,\Delta j}=0$) or signal-dependent QE ($\suma=0$). Under these assumptions:
\begin{eqnarray}
{\rm Err}[\Delta\bar V] &=& [K^2a'+KK']_{0,0} \frac{(I\Delta t)^2}{g^2}(\mu^2-\mu'{^2}),
\nonumber \\
{\rm Err}[\bar C_{\rm H}] &=& \frac{[K^2a'+KK']_{1,0} + [K^2a'+KK']_{-1,0}}{2} \frac{(I\Delta t)^2}{g^2}\mu^2,
\nonumber \\
{\rm Err}[\bar C_{\rm V}] &=& \frac{[K^2a'+KK']_{0,1} + [K^2a'+KK']_{0,-1}}{2} \frac{(I\Delta t)^2}{g^2}\mu^2,
~~{\rm and}
\nonumber \\
{\rm Err}[c_0] &=& {\rm Err}[c_1] = 0.
\label{eq:err-bfe}
\end{eqnarray}
One must iteratively perform the advanced characterization computation in this section and solve for the $[K^2a'+KK']$ kernel via the procedure in \S\ref{ss:char-ipnl} until all parameters are converged.

A similar approach is used for the {\tt nlipc} mode, where the IPNL kernel is attributed entirely to NL-IPC instead of the BFE. In this case:
\begin{eqnarray}
{\rm Err}[\Delta\bar V] &=& [K^2a'+KK']_{0,0} \frac{(I\Delta t)^2}{g^2}\left[ \mu(a+\mu) -\mu'(a+\mu') + \frac{d-a-\mu}{2}(\mu-\mu') \right],
\nonumber \\
{\rm Err}[\bar C_{\rm H}] &=& \frac{[K^2a'+KK']_{1,0} + [K^2a'+KK']_{-1,0}}{2} \frac{(I\Delta t)^2}{g^2}\mu \frac{d+a+\mu}{2},
\nonumber \\
{\rm Err}[\bar C_{\rm V}] &=& \frac{[K^2a'+KK']_{0,1} + [K^2a'+KK']_{0,-1}}{2} \frac{(I\Delta t)^2}{g^2}\mu \frac{d+a+\mu}{2},
~~{\rm and}
\nonumber \\
{\rm Err}[c_0] &=& {\rm Err}[c_1] = 0.
\label{eq:err-nlipc}
\end{eqnarray}

\subsection{Characterization of simulated detector data}
\begin{figure}[t!]
\includegraphics[width=6.2in]{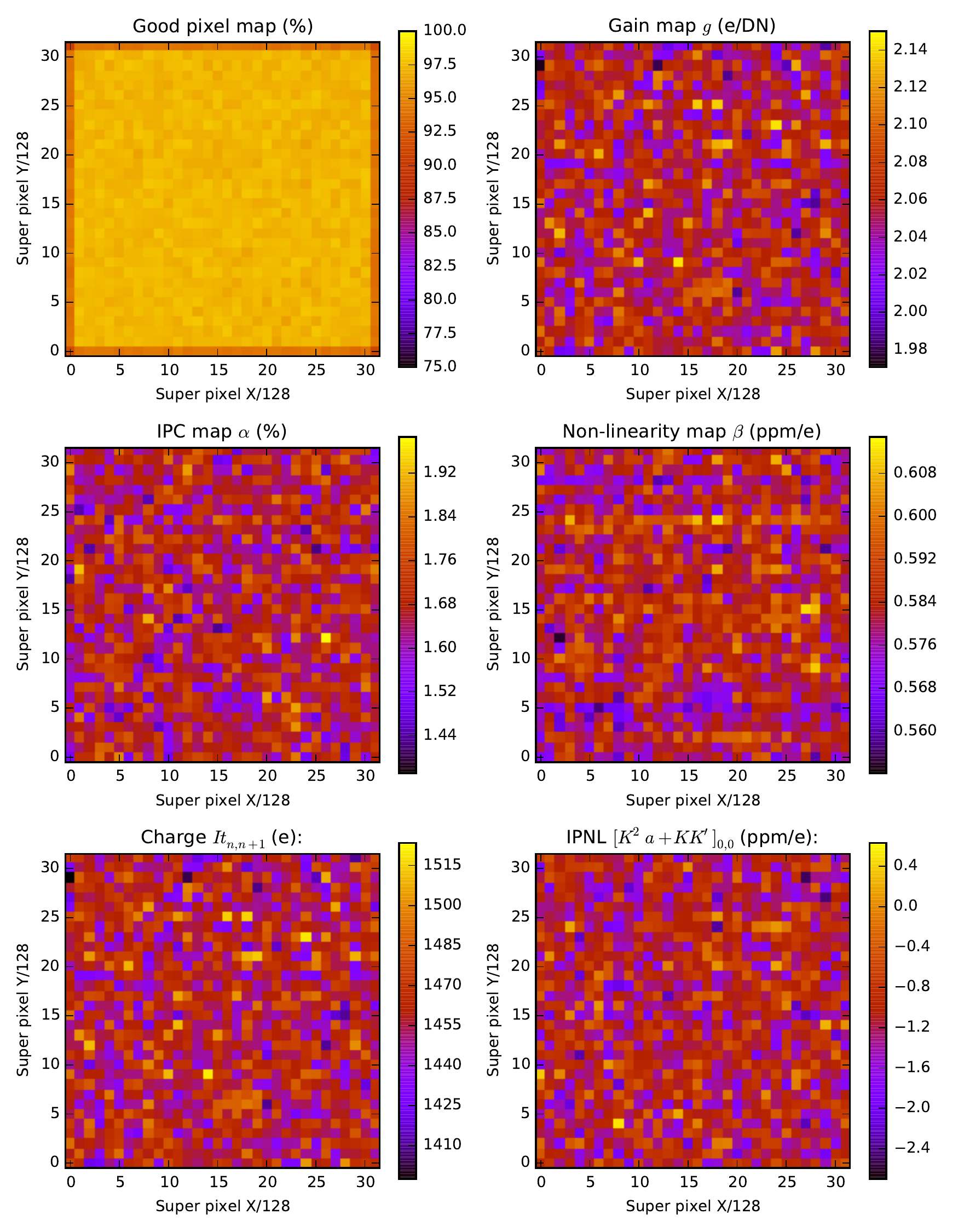}
\caption{\label{fig:advcharsim} Advanced characterization of 3 pairs of simulated flats and darks.}
\end{figure}

In Fig.~\ref{fig:advcharsim}, we show the results of applying the aforementioned advanced characterization steps to pairs of simulated flats and darks using the specifications described in \S~\ref{sec:sim}.  Mean quantities over $N_{good}$ good super-pixels and their statistical errors are provided in Table~\ref{tab:sim_summary}.  The latter values are computed as standard deviations on the mean of the $N_{good}$ super-pixels.

Table~\ref{tab:sim_summary} contains the values of the recovered BFE coefficients obtained after iterative application of the advanced characterization described in this section and the method described in \S~\ref{ss:char-ipnl} (labeled `Method 1').  The time frames used for our fiducial scheme are 3, 11, 13, and 21.  \texttt{solid-waffle} solves for $[K^2a'+KK']$, which reduces to $[K^2a']$ since $K'=0$ in the simulations. $[K^2a']$ values are provided as symmetrical averages for stacks of 3 and 10 simulated flats and compared against the simulation input, where the input $a'$ has been convolved with the input $K^{2}$ (auto-convolution of $K$) to get values comparable to what is actually measured in the correlation analysis. In the central value at zero-lag, $[K^2a']_{0,0}$, we can see there is a bias of 0.1398 ppm/e for the 10 flat stack relative to the input into the simulation (12.1\% bias compared to the input value).  We compute the Method 1 BFE coefficients for two alternative time intervals; the first uses time intervals of half the duration of the fiducial scheme and results in a bias of 11.2\% in the zero-lag coefficient, while the second uses time intervals of twice the fiducial duration and results in a bias of 20.5\%.  We note that the changes to $\beta_{\rm ramp}$ in these alternative time setups are much less than a percent.

We have also run the simulation with only BFE and no IPC and no classical non-linearity.  In the fiducial 3, 11, 13, 21 time frame analysis setup for 10 simulated flats and darks, we obtain $[K^2a']_{0,0}=-1.3225 \pm 0.0077$ (stat) ppm/e, which is biased compared to the input value of -1.3720 ppm/e by 3.6\%.  In this setup, the correct charge per time slice, gain, $\alpha$ and $\beta$ are consistent with the input values (where the latter two are consistent with 0).  We suggest the likely source of bias in the BFE coefficients extracted from the simulations is due to exclusion of higher order terms in the interactions among the BFE, IPC, and classical non-linearity, and we will revisit this investigation in future work. Note that such an investigation of higher-order effects has recently been completed for CCDs \citep{2019arXiv190508677A}.

\begin{table}[]
\scriptsize  
    \centering
    \begin{tabular}{llccccccc}
    \hline\hline
     Quantity     &Units  &\multicolumn{3}{c}{Flat type, number} &\multicolumn{3}{c}{Uncert} &Notes \\
      & & sim,n3 &sim,n10 &truth &stat.(3) &stat.(10) &sys.(3) & \\
      \hline
     Charge, $It_{n,n+1}$    &ke   &1.4607  &1.4615  &1.4604 &0.0006 &0.0003 &\\
     Gain $g$               &e/DN  &2.0606  &2.0620  &2.0600 &0.0008 &0.0004 &\\
     IPC $\alpha$           &\%  &1.6764  &1.6793  &1.6900 &0.0055 &0.0025  & \\
     IPC $\alpha_{\rm H}$&\%  &1.6809  &1.6806  &1.6900 &0.0039  &0.0018 &  \\
     IPC $\alpha_{\rm V}$&\%  &1.6720  &1.6779  &1.6900 &0.0038  &0.0018 &  \\
     IPC $\alpha_{\rm D}$&\%  &-0.0002  &-0.0021  &0.0000 &0.0027  &0.0012 &  \\
     Non-linearity $\beta_{\rm ramp}$  &ppm/e  &0.5835  &0.5782  &0.5800 &0.0003 &0.0001 &0.0091  \\
     \hline
    \multicolumn{8}{c}{Alternative intervals} \\
    Non-linearity $\beta_{\rm ramp}$  &ppm/e &0.5862 &0.5794 &0.5800 &0.0006 &0.0003 &0.0191 & Frames 3,7,9 \\
    Non-linearity $\beta_{\rm ramp}$  &ppm/e &0.5806 &0.5801 &0.5800 &0.0002 &0.0001 &0.0052 &Frames 3,19,21 \\
    \hline
    \multicolumn{8}{c}{Non-overlapping correlation function (Method 1)} \\
    \multicolumn{8}{c}{BFE Coefficients - frames 3,11,13,21, baseline-corrected} \\
    $[K^{2}a^{\prime}]_{0,0}$ &ppm/e &-1.0373  &-1.0192 &-1.1590 &0.0145  &0.0064 &0.0103 &Central pixel \\
    $[K^{2}a^{\prime}]_{<1,0>}$ &ppm/e &0.1838  &0.1980 &0.2034 &0.0073  &0.0033 & &Nearest neighbor \\
    $[K^{2}a^{\prime}]_{<1,1>}$ &ppm/e &0.0362  &0.0428 &0.0505 &0.0072  &0.0032 & &Diagonal \\
    $[K^{2}a^{\prime}]_{<2,0>}$ &ppm/e &0.0155  &0.0133 &0.0120 &0.0074  &0.0032 & & \\
    $[K^{2}a^{\prime}]_{<2,1>}$ &ppm/e &0.0049  &0.0010 &0.0027 &0.0052  &0.0023 & & \\
    $[K^{2}a^{\prime}]_{<2,2>}$ &ppm/e &0.0271  &0.0179 &0.0185 &0.0075  &0.0033 & & \\

    \multicolumn{8}{c}{BFE Coefficients - frames 3,7,9,13 baseline-corrected} \\
    $[K^{2}a^{\prime}]_{0,0}$ &ppm/e &-1.0400 &-1.0293 &-1.1590 &0.0288 &0.0130 &0.0216 &Central pixel \\
    $[K^{2}a^{\prime}]_{<1,0>}$ &ppm/e &0.2381 &0.2195 &0.2034 &0.0152 &0.0066 & &Nearest neighbor \\
    $[K^{2}a^{\prime}]_{<1,1>}$ &ppm/e &0.0479 &0.0392 &0.0505 &0.0151 &0.0066 & &Diagonal \\
    \multicolumn{8}{c}{BFE Coefficients - frames 3,19,21,37 baseline-corrected} \\
    $[K^{2}a^{\prime}]_{0,0}$ &ppm/e &-0.9156 &-0.9214 &-1.1590 &0.0068 &0.0031 &0.0059 &Central pixel \\
    $[K^{2}a^{\prime}]_{<1,0>}$ &ppm/e &0.1850 &0.1818 &0.2034 &0.0034 &0.0015 & &Nearest neighbor \\
    $[K^{2}a^{\prime}]_{<1,1>}$ &ppm/e &0.0422 &0.0450 &0.0505 &0.0035 &0.0016 & &Diagonal \\
    \hline
    \multicolumn{8}{c}{Mean-variance relation (Method 2)} \\
    $\hat{a}_{0,0,M2}$ &ppm/e &-1.3120 &-1.3513 &-1.3720 &0.0383 &0.0144 &0.0273 & \\
    $\beta-4(1+3\alpha)\alpha^{\prime}$ &ppm/e &0.5613 &0.5677 &0.5800 &0.0218 &0.0079 & & \\
    $\sum_{a}-8(1+3\alpha)\alpha^{\prime}$ &ppm/e &-0.0445 &-0.0211 &0.0000 &0.0436 &0.0158 &0.0182 & \\
    \hline
    \multicolumn{8}{c}{Adjacent pixel correlations (Method 3)} \\
    $[K^{2}a^{\prime}]_{<1,0>}-\alpha\sum_{a}$ &ppm/e &0.1816 &0.1855 &0.2034 &0.0072 &0.0032 & &
    \end{tabular}
    \caption{Averaged results for the simulations, based on stacks of flat ramps.  These values were obtained with advanced characterization with ncycle=3.}
    \label{tab:sim_summary}
\end{table}

\subsection{Raw gain and equal-interval correlation tests}

The above techniques enable us to correct the measured properties (gain, IPC, and non-linearity) for the IPNL -- {\em if} we know whether to interpret the non-overlapping correlation function as BFE, NL-IPC, or a mixture of the two. Fortunately, the flat field auto-correlations carry enough information to distinguish the sources of IPNL. We cannot do this based on the non-overlapping correlation function, since in that case both BFE and NL-IPC scale as $\propto t_{ab}t_{cd}$, but we can use the scalings of the raw gain $\hat g^{\rm raw}_{abad}$ and the adjacent-pixel correlations $C_{adad}(\langle 1,0 \rangle)$ as a function of which intervals in the flat field are taken.

\subsubsection{Raw gain vs. interval duration}

In this case, the key observable is the mean-variance slope, in the form $\hat g^{\rm raw}_{abad}$. From Eq.~(\ref{eq:MV-slope-abd}), one sees that there should be two time dependences: one that depends on the start time $t_a$ and contains only the classical non-linearity $\beta$, and one that depends on the duration pattern ($t_{ab}$ and $t_{ad}$) and depends on both $\beta$ and ${a}_{0,0}$. In this section, we consider the first dependence. We fix $t_a$ and fit a linear equation of the form:
\begin{equation}
\ln \hat g^{\rm raw}_{abad} = C_0 + C_1 I(t_{ad}+t_{ab}),
\label{eq:log-slope}
\end{equation}
where $C_0$ is the intercept and $C_1$ is the slope.\footnote{An alternative, which we tried first, is to do a linear fit $\hat g^{\rm raw}_{abd} = B_0 + B_1 I(t_{ad}+t_{ab})$, and use the slope-to-intercept ratio $B_1/B_0$. This procedure is not stable because the intercept $B_0$ is obtained by extrapolating to $t_{ad}+t_{ab}=0$. There is therefore a strong anti-correlation between the slope and intercept, which results in a noise bias: $B_1/B_0$ is biased upward by an amount $-{\rm Cov}(B_0,B_1)/B_0^2$. The amount of bias increases as subsets of the data are used. The formulation of Eq.~(\ref{eq:log-slope}) avoids this problem.} From Eq.~(\ref{eq:MV-slope-abd}), we interpret the slope  as
\begin{equation}
C_1 = 3\beta -(1+8\alpha) [K^2a]_{0,0} + 8(1+3\alpha)\alpha'
= \left\{\begin{array}{lll}
3\beta_{\rm r} & & {\tt none} \\
3\beta_{\rm r} - (1+8\alpha) [K^2a']_{0,0} & & {\tt bfe} \\
3\beta_{\rm r} - 2(1+8\alpha) [KK']_{0,0} & & {\tt nlipc}
\end{array}\right.,
\label{eq:Method-2a}
\end{equation}
where the three possibilities on the right are for no IPNL ({\tt none}), and for the cases where the IPNL is pure BFE ({\tt bfe}) or pure NL-IPC ({\tt nlipc}). If there is a measurement of $[K^2a'+KK']$ from the non-overlapping correlation function, then Eq.~(\ref{eq:Method-2a}) can be used to test these hypotheses about its origin.

\begin{figure}[ht]%
    \centering
    {\includegraphics[width=0.45\textwidth]{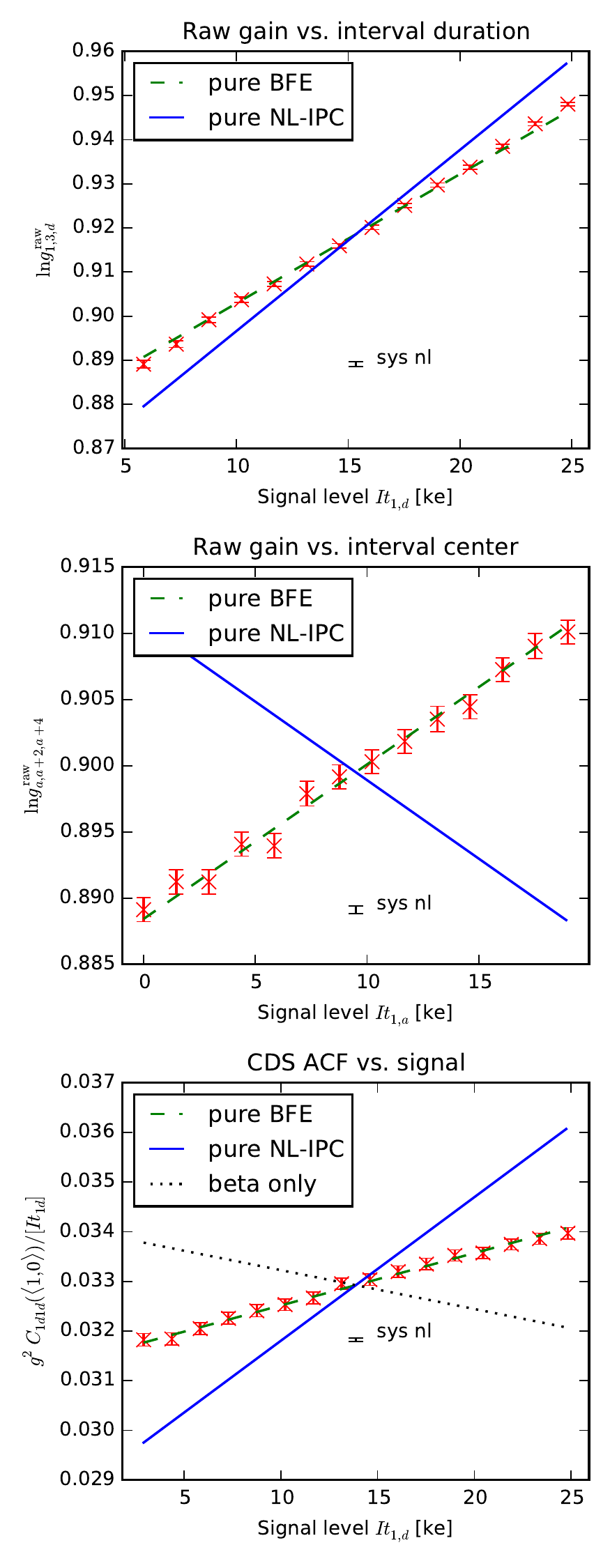}}%
    \caption{Visual comparison of BFE predictions from Method 1 vs measurements from Methods 2 and 3 for simulated detector data (3 flats).}%
    \label{fig:sim_bfe_m23}%
\end{figure}

We compute the raw gain for frame triplets from [1,3,5], [1,3,6],...,[1,5,18] as a function of the signal level accumulated between the first time slice and the time slice $d=5...18$ for the simulated detector data. The top panel of Figure~\ref{fig:sim_bfe_m23} visualizes the results of this test.  Each data point is a mean over all super-pixels, with an error bar based on the error on the mean.  The dashed line is the \texttt{bfe} interpretation of quantities from Method 1, as given by Eq.~\ref{eq:Method-2a}, and the solid line is the \texttt{nlipc} interpretation.  These lines are plotted such that the central values pass through the center of the measurements.  The simulated data agree firmly with the \texttt{bfe} slope, as is expected.

In each panel of Figure~\ref{fig:sim_bfe_m23}, we also show a systematic error related to the modeling of the non-linearity (``sys nl''). This is based on fitting a 5th order polynomial to the median signal levels in the detector. For both this 5th order curve and the quadratic ($\beta$) model, we computed the expected raw logarithmic gain $\ln g^{\rm raw}_{a,b,d}$ for Poisson statistics, compute the difference, and plot an error bar showing the peak$-$valley range. For the case of these simulated data this systematic is negligible, however we include it in anticipation of the analysis of the real data in Paper II where there may be deviations of the classical non-linearity from the $\beta$ model.

We can also make an estimate of the zero-lag BFE coefficient by re-arranging the left part of Eq.~\ref{eq:Method-2a} and substituting $\beta_{\rm r}=\beta-\frac{1}{2}\suma$:
\begin{eqnarray}
\hat{a}_{0,0,\rm M2} &\equiv& a_{0,0}+8\alpha a_{<1,0>}-\frac{3}{2}\suma -8(1+3\alpha)\alpha^{\prime} = 3\beta_{\rm r}-C_{1}
\label{eq:Method-2a_a00}
\end{eqnarray}
Since we did not include non-linear IPC in the simulations, Eq.~\ref{eq:Method-2a_a00} simplifies to $a_{0,0}+8\alpha a_{<1,0>}$. For the 10 simulated flats, $\hat{a}_{0,0,M2}=-1.3513 \pm 0.0144$ (stat) ppm/e.  The input value is -1.3341 ppm/e, so these values agree to within 1.2$\sigma$.

\subsubsection{Raw gain vs. interval center}

A similar test can be carried out by measuring how the raw gain $\hat g^{\rm raw}_{abad}$ varies with $t_a$ as $t_{ab}$ and $t_{ad}$ are held fixed. We fit:
\begin{equation}
\ln \hat g^{\rm raw}_{abad} = C'_0 + C'_1 I t_a.
\label{eq:log-slope.2}
\end{equation}
In this case, we see that one should have
\begin{equation}
C'_1 = 2\beta - 8(1+3\alpha)\alpha'
= \left\{\begin{array}{lll}
2\beta_{\rm r} & & {\tt none} \\
2\beta_{\rm r} & & {\tt bfe} \\
2\beta_{\rm r} - 2(1+8\alpha) [KK']_{0,0} & & {\tt nlipc}
\end{array}\right..
\label{eq:Method-2b}
\end{equation}
Note that the slope $C'_1$ has no sensitivity to the BFE -- the {\tt none} and {\tt bfe} cases give identical predictions. It is however sensitive to NL-IPC.

We compute the raw gain for frame triplets from [1,3,5], [2,4,6],$...$,[14,16,18], as a function of the signal level accumulated between the first time slice and the time slice $a=1,...,14$ for the simulated detector data.  The middle panel of Figure~\ref{fig:sim_bfe_m23} visualizes the results of this test, showing that the simulated data are again consistent with the \texttt{bfe} slope.

Re-writing Eq.~\ref{eq:Method-2b} and using the fact that $\alpha'=0$, we can also compute $\beta=\frac{1}{2} C_{1}^{\prime}$ and $\suma=C_{1}^{\prime}-2\beta_{\rm r}$.  $\beta=0.5677 \pm 0.0079$ ppm/e, which is very close to the input value of 0.58 ppm/e (within 1.6$\sigma$).  Likewise, $\suma=-0.0211 \pm 0.0158$ is very close to the expected value of 0.

\subsubsection{CDS autocorrelation vs. signal}

This method uses the equal-interval correlation function in adjacent pixels, Eq.~(\ref{eq:auto-nearest}). Once the preliminary characterization of the detector has been performed, we may fix the starting time $t_a$ and fit the combination $g^2C_{abab}(\pm 1,0)/(It_{ab})$ as a function of $t_{ab}$, i.e., we fit
\begin{equation}
\frac{g^2}{It_{ab}}C_{abab}(\langle \pm 1,0 \rangle) = C''_{0} + C''_{1} It_{ab}.
\label{eq:M3fit}
\end{equation}
The slope is given by
\begin{equation}
C''_{1} =  -8\alpha\beta + \alpha \suma + [K^2a]_{\langle 1,0\rangle} +2 [KK']_{\langle 1,0\rangle}
= \left\{\begin{array}{lll}
-8\alpha\beta_{\rm r} & & {\tt none} \\
-8\alpha\beta_{\rm r} + [K^2a']_{\langle 1,0\rangle} & & {\tt bfe} \\
-8\alpha\beta_{\rm r} + 2[KK']_{\langle 1,0\rangle} & & {\tt nlipc} \end{array}\right..
\label{eq:Method-3}
\end{equation}
Adding $8\alpha_{\rm H}\beta_{\rm r}$ to the left hand part of Eq.~\ref{eq:Method-3} gives
\begin{equation}
    C''_{1} +8\alpha\beta_{\rm r} = [K^2a]_{\langle 1,0\rangle} +2[KK']_{\langle 1,0\rangle} - 3\alpha \suma = [K^2a'+2KK']_{\langle 1,0\rangle}-\alpha \suma .
\end{equation}

We measure the IPC via basic characterization of frame triplets from [1,2,3], [1,2,4],..., [1,2,18], and CDS auto-correlations for [frame 3 - frame 1], [frame 4 - frame 1],..., [frame 18 - frame 1]. The bottom panel of Figure~\ref{fig:sim_bfe_m23} visualizes the results of this test on the simulated detector data, which are consistent with the \texttt{bfe} interpretation.

We expect that $[K^2a'+2KK']_{\langle 1,0\rangle}-\alpha \suma$ simplifies to $[K^2a']_{\langle 1,0\rangle}$ for the simulated data.  This value is $0.1855 \pm 0.0032$ ppm/e and can be compared with the value obtained from Method 1 of $[K^2a']_{\langle 1,0\rangle}=0.1980 \pm 0.0033$ and the input value of 0.2034 ppm/e ($\sim$9\% difference between the input and the value obtained with the CDS autocorrelation method).

\section{Discussion}
\label{sec:discussion}

In this paper, we present formalism to connect flat field correlations to various detector effects in infrared detector arrays, including non-linear effects such as the BFE and NL-IPC.  This formalism is built up through first considering the Poisson statistics in a perfect detector and then including contributions from the IPC kernel, classical non-linearity, BFE, and NL-IPC.  In the expression for the combined cross-correlation of two CDS images (sampled at time frames $a, b, c, d$), we consider the leading order interactions, namely $\alpha$, $\alpha^2$, $\beta$, $\alpha\beta$, $a$, $\alpha a$, $\alpha'$, and $\alpha\alpha'$.  We discuss two special cases of the combined correlation function: the non-overlapping correlation function ($a<b<c<d$), which has the most sensitivity to the inter-pixel non-linear effects, but cannot by itself distinguish between the BFE and NL-IPC; the equal-interval correlation function ($a=c<b=d$), which is the auto-correlation of a CDS image and is most similar to the flat field statistics available for CCDs.  We also discuss features of the raw gain for the case of ($a=c<b<d$), which provides a means of distinguishing between the BFE and NL-IPC interpretations through the different behaviors of these mechanisms as a function of time.

We describe a procedure for characterizing detector arrays and extracting measurements of the IPNL.  This involves constructing CDS images, performing a reference pixel subtraction, computing the raw gain, IPC, correlations (in the horizontal, vertical, and diagonal directions), and ramp curvature; we use these to solve for $g$, $\alpha_{\rm H}$, $\alpha_{\rm V}$, $I$, and $\beta$.  We show how to use the non-overlapping correlation function to obtain the IPNL and also how to apply an iterative scheme to correct the $g$, $\alpha_{\rm H}$, $\alpha_{\rm V}$, $I$, and $\beta$ for residual biases imprinted by the IPNL.

We validate our methodology on simulated flat fields, which are constructed to imitate characteristics ($g$, $\alpha_{\rm H}$, $\alpha_{\rm V}$, $I$, and $\beta$) of the real detector array tested in Paper II.  For this first investigation, we input a BFE kernel (but no NL-IPC).  We extract parameters that match the inputs with high accuracy, except for the BFE kernel, for which we obtain a zero-lag component which is biased by 12\%.  We also show that the raw gain and equal-interval correlation function interpretation tests are successful in distinguishing between the BFE and NL-IPC as the underlying mechanism for the IPNL in the simulations.  Given the success in obtaining equivalent inputs and outputs of the other key parameters, namely $\beta$ and $\alpha$, we suggest the 12\% bias in the extracted BFE kernel could likely be explained by unaccounted interactions at higher orders that were dropped in the approximations used in this work.  The impact of these higher order terms is under investigation and will be addressed in future work.

\section*{Acknowledgements}

We thank the Detector Characterization Laboratory personnel, Yiting Wen, Bob Hill, and Bernie Rauscher at NASA Goddard Space Flight Center for their efforts enabling the existence and access to the data analyzed in this series of papers, and we thank Chaz Shapiro, Andr\'es Plazas, and Eric Huff for helpful discussions. We thank Jay Anderson and Arielle Bertrou-Cantou for useful presentations to the Detector Working Group on their analyses of non-linearities in the HST/WFC3-IR and Euclid H2RG detectors. We thank the anonymous referee for helpful suggestions that improved the clarity of this paper.  We are also grateful for the use of \citet{OhioSupercomputerCenter1987} for computing the results in this work. AC and CMH acknowledge support from NASA grant 15-WFIRST15-0008. During the preparation of this work, CMH has also been supported by the Simons Foundation and the US Department of Energy.  Software: Astropy \citep{2013A&A...558A..33A,2018AJ....156..123A}, fitsio \citep{fitsio}, Matplotlib \citep{Hunter:2007}, NumPy \citep{numpy}, SciPy \citep{scipy}

\appendix

\section{Clipping correction to the covariance}
\label{app:covar-clip}

This appendix considers the correction to the covariance matrix of two jointly Gaussian distributed variables, $X$ and $Y$, when those distributions are clipped. We are interested in the parameter $f_{\rm corr}$ defined by
\begin{equation}
{\rm Cov}(X,Y)|_{\rm clipped} = f_{\rm corr} {\rm Cov}(X,Y)|_{\rm true}.
\label{eq:fcorr}
\end{equation}
We assume that a fraction $\epsilon$ of the data are clipped from both the top and the bottom of the distribution in $X$ and $Y$; if $X$ and $Y$ were independent, this would mean that a fraction $(1-2\epsilon)^2$ of the data points survive the clipping, but the fraction that survives may be larger if $X$ and $Y$ are covariant.

The determination of $f_{\rm corr}$ is invariant to linear rescaling of $X$ and $Y$, so without loss of generality, we assume that $X$ and $Y$ both have mean 0 and variance 1. Their ``true'' covariance is then the correlation coefficient $\rho$. The clipping is equivalent to the restriction of the data at $|X|, |Y|<\xi$, where
\begin{equation}
{\cal P}(\xi) \equiv \int_{-\infty}^\xi \frac1{\sqrt{2\pi}} e^{-z^2/2}\,dz = 1-\epsilon.
\end{equation}
Since the clipped distribution still has $\langle X\rangle = \langle Y \rangle = 0$ by symmetry, we are interested in the mean value of $XY$:
\begin{equation}
{\rm Cov}(X,Y)|_{\rm clipped} = \frac{\int_{-\xi}^\xi dX \int_{-\xi}^\xi dY\, p(X,Y) XY}{\int_{-\xi}^\xi dX \int_{-\xi}^\xi dY\, p(X,Y)},
\end{equation}
where the denominator is the survival probability of a data point $(X,Y)$, and the probability distribution is
\begin{equation}
p(X,Y) = \frac1{2\pi(1-\rho^2)} e^{-(X^2+Y^2-2\rho XY)/2(1-\rho^2)}.
\end{equation}
The covariance can be expanded in a power series in $\rho$; the leading term is
\begin{equation}
{\rm Cov}(X,Y)|_{\rm clipped} = \left[ 1 - \sqrt{\frac2\pi} \frac{\xi e^{-\xi^2/2}}{1-2\epsilon} \right]^2 \rho + {\cal O}(\rho^3),
\end{equation}
so that
\begin{equation}
f_{\rm corr} = \left( 1 - \sqrt{\frac2\pi} \frac{\xi e^{-\xi^2/2}}{1-2\epsilon} \right)^2 + {\cal O}(\rho^2).
\label{eq:fcorr-0}
\end{equation}
The clipped covariances used in this paper to measure IPC are corrected using the leading constant term in $f_{\rm corr}$. The correction factor should converge to 1 as $\epsilon\rightarrow 0$; this is easily verified.

Note that the ``correction'' is not small: for $\epsilon=0.01$ (i.e.\ clipping the top 1\%\ and bottom 1\%\ of the distribution) we have $f_{\rm corr} = 0.7629$. If one clips more of the distribution, the correction becomes enormous: at $\epsilon=0.025$ we have $f_{\rm corr} = 0.5758$. Going the other way, even for $\epsilon = 10^{-3}$, the correction is $f_{\rm cor} = 0.9587$.

\bibliographystyle{aasjournal}
\bibliography{main}

\end{document}